\documentclass[ALICE,manyauthors]{cernphprep}
    
\usepackage[comma,square,numbers,sort&compress]{natbib}

\usepackage{lineno}
\usepackage{xspace}
\usepackage{hyperref}
\usepackage{bm}
\usepackage{newtxtext,newtxmath}
\usepackage{upgreek}

\usepackage[T1]{fontenc}
\usepackage{orcidlink}

\usepackage{xcolor}
\begin{document}
%

\newcommand{\pp}           {pp\xspace}
\newcommand{\ppbar}        {\mbox{$\mathrm {p\overline{p}}$}\xspace}
\newcommand{\XeXe}         {\mbox{Xe--Xe}\xspace}
\newcommand{\PbPb}         {\mbox{Pb--Pb}\xspace}
\newcommand{\pA}           {\mbox{pA}\xspace}
\newcommand{\pPb}          {\mbox{p--Pb}\xspace}
\newcommand{\AuAu}         {\mbox{Au--Au}\xspace}
\newcommand{\dAu}          {\mbox{d--Au}\xspace}

\newcommand{\s}            {\ensuremath{\sqrt{s}}\xspace}
\newcommand{\snn}          {\ensuremath{\sqrt{s_{\mathrm{NN}}}}\xspace}
\newcommand{\pt}           {\ensuremath{p_{\rm T}}\xspace}
\newcommand{\meanpt}       {$\langle p_{\mathrm{T}}\rangle$\xspace}
\newcommand{\ycms}         {\ensuremath{y_{\rm CMS}}\xspace}
\newcommand{\ylab}         {\ensuremath{y_{\rm lab}}\xspace}
\newcommand{\etarange}[1]  {\mbox{$\left | \eta \right |~<~#1$}}
\newcommand{\yrange}[1]    {\mbox{$\left | y \right |~<~#1$}}
\newcommand{\dndy}         {\ensuremath{\mathrm{d}N_\mathrm{ch}/\mathrm{d}y}\xspace}
\newcommand{\dndeta}       {\ensuremath{\mathrm{d}N_\mathrm{ch}/\mathrm{d}\eta}\xspace}
\newcommand{\avdndeta}     {\ensuremath{\langle\dndeta\rangle}\xspace}
\newcommand{\dNdy}         {\ensuremath{\mathrm{d}N_\mathrm{ch}/\mathrm{d}y}\xspace}
\newcommand{\Npart}        {\ensuremath{N_\mathrm{part}}\xspace}
\newcommand{\Ncoll}        {\ensuremath{N_\mathrm{coll}}\xspace}
\newcommand{\dEdx}         {\ensuremath{\textrm{d}E/\textrm{d}x}\xspace}
\newcommand{\RpPb}         {\ensuremath{R_{\rm pPb}}\xspace}
\newcommand{\Minv}         {\ensuremath{M_{\uppi^{+}\uppi^{-}}}\xspace}
\newcommand{\ctaueff}{\mbox{$\left\langle c\tau_{\mathrm{res}}^{\mathrm{eff}} \right\rangle$}\xspace}
\newcommand{\masseff}{\mbox{$\left\langle m_{\mathrm{res}}^{\mathrm{eff}} \right\rangle$}\xspace}

\newcommand{\nineH}           {$\sqrt{s}~=~0.9$~Te\kern-.1emV\xspace}
\newcommand{\seven}           {$\sqrt{s}~=~7$~Te\kern-.1emV\xspace}
\newcommand{\eight}            {$\sqrt{s}~=~8$~Te\kern-.1emV\xspace}
\newcommand{\thirteen}        {$\sqrt{s}~=~13$~Te\kern-.1emV\xspace}
\newcommand{\thirteensix}    {$\sqrt{s}~=~13.6$~Te\kern-.1emV\xspace}
\newcommand{\fourteen}        {$\sqrt{s}~=~14$~Te\kern-.1emV\xspace}
\newcommand{\twoH}            {$\sqrt{s}~=~0.2$~Te\kern-.1emV\xspace}
\newcommand{\twosevensix}  {$\sqrt{s}~=~2.76$~Te\kern-.1emV\xspace}
\newcommand{\five}               {$\sqrt{s}~=~5.02$~Te\kern-.1emV\xspace}
\newcommand{\fivethirtysix}   {$\sqrt{s}~=~5.36$~Te\kern-.1emV\xspace}
\newcommand{\fivefive}         {$\sqrt{s}~=~5.5$~Te\kern-.1emV\xspace}
\newcommand{\twosevensixnn}{$\sqrt{s_{\mathrm{NN}}}~=~2.76$~Te\kern-.1emV\xspace}
\newcommand{\fivenn}           {$\sqrt{s_{\mathrm{NN}}}~=~5.02$~Te\kern-.1emV\xspace}
\newcommand{\fivethirtysixnn}   {$\sqrt{s_{\mathrm{NN}}}~=~5.36$~Te\kern-.1emV\xspace}
\newcommand{\fivefivenn}         {$\sqrt{s_{\mathrm{NN}}}~=~5.5$~Te\kern-.1emV\xspace}
\newcommand{\LT}           {L{\'e}vy-Tsallis\xspace}
\newcommand{\GeVc}         {Ge\kern-.1emV/$c$\xspace}
\newcommand{\MeVc}         {Me\kern-.1emV/$c$\xspace}
\newcommand{\TeV}          {Te\kern-.1emV\xspace}
\newcommand{\GeV}          {Ge\kern-.1emV\xspace}
\newcommand{\MeV}          {Me\kern-.1emV\xspace}
\newcommand{\tev}          {Te\kern-.1emV\xspace}
\newcommand{\gev}          {\rm Ge\kern-.1emV\xspace}
\newcommand{\mev}          {\rm Me\kern-.1emV\xspace}
\newcommand{\GeVmass}      {Ge\kern-.2emV/$c^2$\xspace}
\newcommand{\MeVmass}      {Me\kern-.2emV/$c^2$\xspace}
\newcommand{\lumi}         {\ensuremath{\mathcal{L}}\xspace}

\newcommand{\ITS}          {\rm{ITS}\xspace}
\newcommand{\TOF}          {\rm{TOF}\xspace}
\newcommand{\ZDC}          {\rm{ZDC}\xspace}
\newcommand{\ZDCs}         {\rm{ZDCs}\xspace}
\newcommand{\ZNA}          {\rm{ZNA}\xspace}
\newcommand{\ZNC}          {\rm{ZNC}\xspace}
\newcommand{\SPD}          {\rm{SPD}\xspace}
\newcommand{\SDD}          {\rm{SDD}\xspace}
\newcommand{\SSD}          {\rm{SSD}\xspace}
\newcommand{\TPC}          {\rm{TPC}\xspace}
\newcommand{\TRD}          {\rm{TRD}\xspace}
\newcommand{\VZERO}        {\rm{V0}\xspace}
\newcommand{\VZEROA}       {\rm{V0A}\xspace}
\newcommand{\VZEROC}       {\rm{V0C}\xspace}
\newcommand{\Vdecay} 	   {\ensuremath{V^{0}}\xspace}

\newcommand{\ee}           {\ensuremath{e^{+}e^{-}}} 
\newcommand{\pip}          {\ensuremath{\uppi^{+}}\xspace}
\newcommand{\pim}          {\ensuremath{\uppi^{-}}\xspace}
\newcommand{\kap}          {\ensuremath{\rm{K}^{+}}\xspace}
\newcommand{\kam}          {\ensuremath{\rm{K}^{-}}\xspace}
\newcommand{\pbar}         {\ensuremath{\rm\overline{p}}\xspace}
\newcommand{\kzero}        {\ensuremath{{\rm K}^{0}_{\rm{S}}}\xspace}
\newcommand{\lmb}          {\ensuremath{\Lambda}\xspace}
\newcommand{\almb}         {\ensuremath{\overline{\Lambda}}\xspace}
\newcommand{\Om}           {\ensuremath{\Omega^-}\xspace}
\newcommand{\Mo}           {\ensuremath{\overline{\Omega}^+}\xspace}
\newcommand{\X}            {\ensuremath{\Xi^-}\xspace}
\newcommand{\Ix}           {\ensuremath{\overline{\Xi}^+}\xspace}
\newcommand{\Xis}          {\ensuremath{\Xi^{\pm}}\xspace}
\newcommand{\Oms}          {\ensuremath{\Omega^{\pm}}\xspace}
\newcommand{\degree}       {\ensuremath{^{\rm o}}\xspace}
\newcommand{\Dzero}     {\ensuremath{\rm D}^{0}\xspace}
\newcommand{\Dplus}     {\ensuremath{\rm D}^{+}\xspace}
\newcommand{\Dstar}     {\ensuremath{\rm D}^{*+}\xspace}
\newcommand{\Dstarwide} {\ensuremath{\rm D}^{*}(2010)^{+}\xspace}
\newcommand{\Lc}        {\ensuremath{\rm \Lambda}_{\rm c}^{+}\xspace}
\newcommand{\Sc}        {\ensuremath{\rm \Sigma}_{\rm c}^{0,++}\xspace}

\newcommand{\aprot}{\ensuremath{\mathrm{\bar{p}}}\,}
\newcommand{\prot}{\ensuremath{\mathrm{p}}\,}
\newcommand{\pP}{\ensuremath{\mathrm {p\mbox{--}p}}\,}
\newcommand{\pXi}{\ensuremath{\mathrm {p\mbox{--}\Xi}}\,}
\newcommand{\pOmega}{\ensuremath{\mathrm {p\mbox{--}\Omega}}\,}
\newcommand{\LXi}{\ensuremath{\mathrm {\Lambda\mbox{--}\Xi}}\,}
\newcommand{\phiP}{\ensuremath{\mathrm {\upphi\mbox{--}p}}\,}
\newcommand{\akP}{\ensuremath{\mathrm {K^-\mbox{--}p}}\,}
\newcommand{\Kp}{\ensuremath{\mathrm {K\mbox{--}p}}\,}
\newcommand{\aKp}{\ensuremath{\mathrm {K^+\mbox{--}p}}\,}
\newcommand{\Kap}{\ensuremath{\mathrm {K^-\mbox{--}\overline{p}}}\,}
\newcommand{\pipi}{\ensuremath{\mathrm {\uppi^+\mbox{--}\uppi^+}}\,}
\newcommand{\apiapi}{\ensuremath{\mathrm {\uppi^-\mbox{--}\uppi^-}}\,}
\newcommand{\spipi}{\ensuremath{\mbox{$\uppi$--$\uppi$}}~}
\newcommand{\pap}{\ensuremath{\mathrm {p\mbox{--}\bar{p}}}\,}
\newcommand{\paL}{\ensuremath{\mathrm {p\mbox{--}\overline{\Lambda}}}\,}
\newcommand{\apL}{\ensuremath{\mathrm {\overline{p}\mbox{--}\Lambda}}\,}
\newcommand{\LaL}{\ensuremath{\mathrm {\Lambda\mbox{--} \overline{\Lambda}}}\,}
\newcommand{\pL}{\ensuremath{\mathrm {p\mbox{--}\Lambda}}\,}
\newcommand{\ApaL}{\ensuremath{\mathrm {\bar{p}\mbox{--}\overline{\Lambda}}}\,}
\newcommand{\BBbar}{\ensuremath{\mathrm {B\mbox{--} \bar{B}}}\,}
\newcommand{\BB}{\ensuremath{\mathrm {B\mbox{--} B}}\,}
\newcommand{\rhop}{\ensuremath{\mathrm {\uprho^{0}\mbox{--} p}}\,}
\newcommand{\rhoap}{\ensuremath{\mathrm {\uprho^{0}\mbox{--} \overline{p}}}\,}

\newcommand{\rs}           {\ensuremath{r^*}\xspace}
\newcommand{\rsv}           {\ensuremath{\vec{r}^*}\xspace}
\newcommand{\rc}           {\ensuremath{r_\mathrm{core}}\xspace}
\newcommand{\rcv}           {\ensuremath{\vec{r}_\mathrm{core}}\xspace}
\newcommand{\rcs}           {\ensuremath{r^*_\mathrm{core}}\xspace}
\newcommand{\rcsv}           {\ensuremath{\vec{r}^*_\mathrm{core}}\xspace}
\newcommand{\ks}           {\ensuremath{k^*}\xspace}
\newcommand{\ksSq}           {\ensuremath{k^{*2}}\xspace}
\newcommand{\ksv}           {\ensuremath{\vec{k}^*}\xspace}
\newcommand{\Sr}           {\ensuremath{S(r)}\xspace}
\newcommand{\Ck}           {\ensuremath{C(k)}\xspace}
\newcommand{\Srs}           {\ensuremath{S(\rs)}\xspace}
\newcommand{\Cks}           {\ensuremath{C(\ks)}\xspace}
\newcommand{\mt}           {\ensuremath{m_{\mathrm{T}}}\xspace}


\begin{titlepage}
\PHyear{2025}       
\PHnumber{179}      
\PHdate{6 August}  

\title{First direct access to  the $\bm{\uprho}^0$p interaction via correlation studies at the LHC}
\ShortTitle{First Direct Access to the $\uprho^0$p Interaction via Correlations Studies at the LHC}   

\Collaboration{ALICE Collaboration\thanks{See Appendix~\ref{app:collab} for the list of collaboration members}}
\ShortAuthor{ALICE Collaboration} 

\begin{abstract}
Direct measurements of the $\uprho^0$p interaction have remained so far elusive, with most insights derived indirectly from photoproduction or low-energy partial wave analyses. This letter presents the first direct observation of the $\uprho^0$p interaction, obtained through two-particle correlations measured in high-multiplicity, ultrarelativistic proton--proton collisions at $\sqrt{s} = 13~\mathrm{TeV}$ by the ALICE Collaboration at the LHC. Two-particle correlation data, analyzed within chiral effective field theory ($\chi$EFT) using a coupled-channel approach and incorporating recent \phiP data, yield a scattering length of $a_{\uprho^0\mathrm{p}} = (-0.46 \pm 0.04) + i(0.20 \pm 0.04)$~fm and constrain coupling strengths of two states identified with the N(1958) and N(1700). These findings emphasize the importance of coupled-channel dynamics and dynamically generated states in understanding the $\uprho^0$p interaction. The results establish a vacuum baseline for extrapolation studies to high densities, contributing to the foundation for chiral symmetry restoration searches, and offer collider-based insights into the QCD spectrum, complementing traditional low-energy approaches. This work marks a significant advance in correlation studies, extending the exploration of interactions to the most short-lived QCD states.

\end{abstract}
\end{titlepage}

\setcounter{page}{2} 


\noindent Direct access to the final-state interaction (FSI) between hadrons and its interpretation within Quantum Chromodynamics (QCD) — recognized as the fundamental theory of the strong interaction — remains a formidable challenge for nuclear physics, primarily due to the intrinsically non-perturbative and confining nature of QCD. By describing interactions directly in terms of the observable low-energy degrees of freedom, represented by hadrons, chiral effective field theories (\(\chi\)EFT) have successfully addressed this difficulty and yielded a wealth of insight while adhering to the fundamental symmetries of QCD, especially the approximate chiral symmetry (CS). The spontaneous breaking of CS, driven by the non-trivial QCD vacuum and characterized by a non-zero expectation value of the chiral quark condensate, is responsible for generating the bulk of the mass of hadronic bound states, endowing the hadron spectrum with structure — a result rigorously derived from QCD sum rules~\cite{PhysRevC.46.R34, doi:10.1142/S0218301310014728, Zschocke2002, PhysRevLett.66.2720}. This connection implies that any modification of the QCD vacuum, such as those induced by a medium (e.g., finite temperature or density), will alter hadron properties like masses and decay widths~\cite{BROCKMANN199640}. These changes are experimentally observable as shifts in spectral functions, providing a unique window into QCD dynamics and hinting at chiral symmetry restoration (CSR) under extreme conditions~\cite{Rapp2000, refId0}. Such a medium can be realized in high-energy heavy-ion collisions, where large energy densities and temperatures are reached, and the system persists for up to 10 fm/$c$~\cite{ALICE:2022wpn}.

Vector mesons (\(\uprho\), \(\upomega\), \(\upphi\)) are ideal probes for studying such modifications. With lifetimes ranging from a few to several tens of fm/$c$, they predominantly decay within the medium, and their dilepton decay channels (\(\rm{e}^+\rm{e}^-\), \(\upmu^+\upmu^-\)), as colour singlets, penetrate the medium without strong interaction. These features make vector mesons especially susceptible to in-medium effects and the \(\uprho\)-meson the ideal probe and primary focus of pertinent experimental efforts. Experiments exploring energy scales from a few GeV to the TeV range, such as those at the CERN-SPS (CERES and NA60)~\cite{Dusling:2006yv, Ruppert:2006hf, CERES:2006wcq, NA60:2006ymb}, HADES~\cite{HADES:2020kce, FRIMAN1997496}, PHENIX~\cite{PHENIX:2008qav, PHENIX:2009gyd, PHENIX:2015vek}, STAR~\cite{STAR:2012dzw, STAR:2013pwb, Huck:2014mfa,  STAR:2015tnn, STAR:2015zal, STAR:2018ldd, STAR:2023wta}, and ALICE~\cite{ALICE:2018ael, ALICE:2023jef}, have extensively investigated dilepton mass spectra for signatures of CSR. Among these, data from STAR and CERN-SPS  (CERES and NA60) exhibit a broadening of the \(\uprho\)-meson spectral function, interpreted as an in-medium effect~\cite{ Dusling:2006yv, Ruppert:2006hf, CERES:2006wcq, NA60:2006ymb, NA60:2008ctj, STAR:2023wta}.  

However, the interpretation of these observations rely on the assumed direct coupling
of the \(\uprho\)-meson to the surrounding nucleons, which drives the in-medium \(\uprho\)-meson self-energy $\Sigma_{\uprho \rm{N}}$~\cite{Rapp:1997ei, Rapp:1999us}, and leads to the excitation of nucleonic resonances in the system. The latter introduces model-dependent uncertainties in the FSI, which complicate the extraction of genuine CSR effects. This interaction is typically modelled using the phenomenological Vector Meson Dominance (VMD) framework~\cite{PhysRevLett.22.981} (see for a short review Ref.~\cite{Schildknecht:2005xr}), which posits that vector mesons interact with photons through their timelike electromagnetic form factor — a quantity constrained by photoproduction data~\cite{Rapp:1997ei, Rapp:1999us} and basis for pertinent calculations~\cite{Rapp:2013nxa, Hohler:2013eba}. Despite its widespread use, the VMD framework lacks unique predictive power. This is evident in the low-energy fixed-target results from HADES, where different VMD implementations yield inconsistent outcomes~\cite{PhysRevC.106.064910, HADES:2022vus}. Moreover, the \(\uprho \rm{N}\) interaction is highly intricate, involving contributions from numerous resonant states~\cite{FRIMAN1997496, Rapp:1999us, HADES:2022vus} and coupled channel effects~\cite{Feijoo:2024bvn}. These complexities, combined with the inherent model dependencies of existing approaches, call for a paradigm shift in the way the \(\uprho \rm{N}\) interaction is studied.

In recent years, driven by advances in understanding the particle production in small systems~\cite{ALICE:2020ibs, ALICE:2023sjd, ALICE:2024Marcel, Mihaylov:2023pyl}, two-particle correlation techniques have been employed to study FSI in ultrarelativistic proton--proton (\pp) collisions for non-identical pairs~\cite{ALICE:2020mfd, ALICE:2019hdt, ALICE:2022yyh, ALICE:2021cpv} including even charmed baryons~\cite{ALICE:2022enj, ALICE:2024bhk}. In this letter, the method is extended to the \rhop system
where a direct measurement of the \(\uprho \rm{N}\) interaction is obtained and interpreted within the framework of \(\chi\)EFT. The main observable of interest is the two-particle correlation function, \Cks~\cite{Lisa:2005dd, 2021ARNPS..71..377F}, measured as the normalised ratio $\Cks~=~\mathcal{N} N_{\mathrm{same}}(\ks)/N_{\mathrm{mixed}}(\ks)$,
where $N_{\mathrm{same}}(\ks)$ and $N_{\mathrm{mixed}}(\ks)$  are the distributions of the relative momentum $(\ks~=~\frac{1}{2}|\Vec{p}_1^*~-~\Vec{p}_2^*|)$ in the same and mixed events, respectively. The latter corresponds to the distribution of uncorrelated pairs and is calculated using the mixed-event technique~\cite{Lisa:2005dd, 2021ARNPS..71..377F}. The relative momentum is evaluated in the rest frame of the particle pair (denoted with an asterisk *). The normalization constant is denoted with $\mathcal{N}$ and ensures the proper asymptotic behaviour in the absence of any non-femtoscopic effect i.e., for large relative
momenta \Cks should converge to unity. The information about the spatio-temporal extension of the particle emitting source as well as the FSI encoded in \Cks is accessed via the Koonin-Pratt equation (see Ref.~\cite{Lisa:2005dd} for a derivation) which reads $\Cks~=~\int \mathrm{d^3}\rs S(\rs) |\uppsi(\rs,\ks)|^{2}$ with $S(\rs)$ being the probability density function of emitting a pair at a relative distance \rs, and $\uppsi(\rs,\ks)$ the two-particle relative wave function, whose form is governed by the FSI and possibly quantum statistics.

The data analyzed in this work were collected by the ALICE Collaboration~\cite{ALICEDetector} at the LHC in \pp collisions at \thirteen during the Run 2 data-taking campaign (2015--2018). Comprehensive descriptions of the ALICE detector setup and its performance are provided in Refs.~\cite{ALICEDetector} and~\cite{ALICEperf}, respectively. The following main sub-detectors are used: the $\VZERO$ detector~\cite{ALICEVZERO}, the Inner Tracking System (ITS)~\cite{ALICEITS}, the Time Projection Chamber (TPC)~\cite{ALICETPC}, and the Time-Of-Flight (TOF) detector~\cite{ALICETOF}. The charged-particle tracking and the primary vertex (PV) reconstruction are performed using the combined track information of the ITS and TPC~\cite{ALICEDetector}, located inside a uniform magnetic field with a maximum of $0.5$~T oriented along the beam axis. The ITS, TPC, and TOF cover the full azimuthal angle and  the pseudorapidity interval \etarange{0.9}. High-multiplicity (HM) events are selected online by applying a threshold on the amplitude of the signal in the V0 detectors — an array of scintillators positioned on each side of the interaction point that cover the pseudorapidity ranges 2.8$~<~\eta~<~$5.1 (V0A) and $-3.7~<~\eta~<~-1.7$ (V0C). The event selection criteria used in this study were already employed in several previous analyses~\cite{ALICE:2020ibs, ALICE:2021cpv}. The Silicon Drift Detectors (SDD) layer of the ITS extends coverage beyond $|\eta|=0.8$ and is used as part of the HM trigger logic. The selected events correspond to the 0.17\% highest multiplicity inelastic \pp collisions with at least one charged particle in the range \etarange{1} (referred to as INEL $>$ 0)~\cite{ALICEDetector,ALICEperf}. The resulting data sample contains events with an average multiplicity of approximately 30 charged particles in the pseudorapidity interval \etarange{0.5}~\cite{ALICE:2020mfd}.
Events with multiple primary vertices, identified from track segments in the two innermost layers of the ITS, are tagged as pile-up and removed from the analyzed sample. A selection of $\pm10$ cm on the difference between the reconstructed PV position along the beam axis ($z$-coordinate) and the nominal interaction point is applied to ensure uniform detector coverage within the pseudorapidity range of \etarange{0.8}, used in this work. In total $10^{9}$ HM pp collision at \thirteen are used as an input to the analysis. 

Charged hadrons are identified using information from the TPC and TOF by applying a threshold on the deviation   between the signal hypothesis for the searched particle and the measured values, expressed in terms of multiples $n_{\upsigma}$ of the detector resolution $\upsigma$. The proton candidate selection follows the strategy outlined in Ref.~\cite{ALICE:2020ibs}. The composition of the selected sample, in terms of primary and secondary particles (e.g., material budget, feed down from weak decays), is studied via fitting Monte Carlo (MC) templates, to the experimentally obtained distance of closest approach in the transverse direction (DCA$_{xy}$) to the PV distributions~\cite{ALICE:2020ibs, ALICE:2023sjd}. These events are simulated using PYTHIA8.2~\cite{PYTHIA} (Monash 2013 Tune), transported through the ALICE detector by GEANT3~\cite{GEANT}, and processed by the reconstruction algorithm~\cite{ALICEDetector}. The resulting sample of proton candidates has a \pt-weighted purity (i.e., the fraction of true protons among selected candidates) of $98\%$ and a \pt-weighted primary fraction of $85\%$~\cite{ ALICE:2024Marcel}. The $\uprho^{0}$ candidates are reconstructed via the hadronic decay channel into two oppositely charged pions $\uprho^{0}\rightarrow\uppi^{+}\uppi^{-}$, analogous to Refs.~\cite{STAR:2003vqj, ALICE:2018qdv}. The pion candidate selection follows the methods outlined in Ref.~\cite{ALICE:2023sjd} and results in a \pt-weighted purity of $98\%$ and a \pt-weighted primary fraction of $85\%$, where the latter is corrected for contributions from long-lived strongly decaying resonances.

\begin{figure}[hbt]
\centering
\includegraphics[width=.5\textwidth]{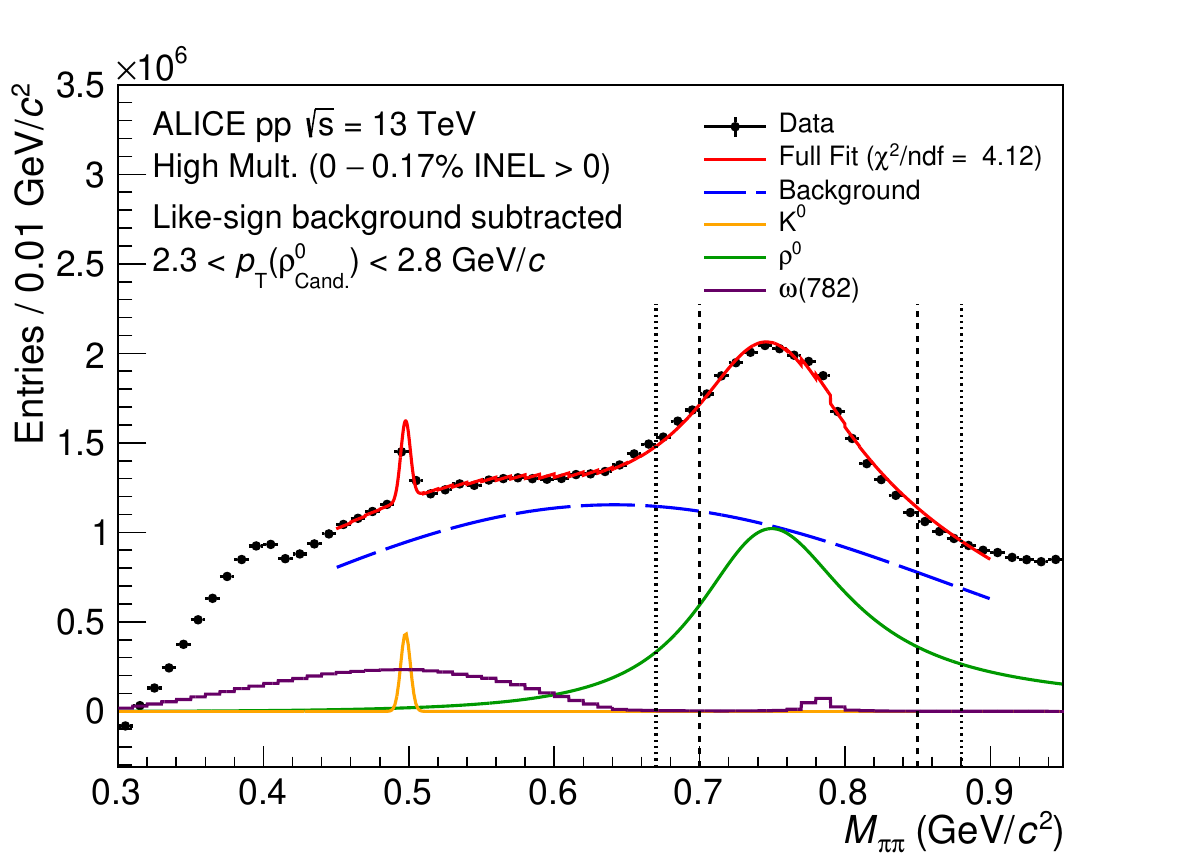}
\caption{Example of the invariant-mass (\Minv) spectrum for a specific transverse momentum interval of the $\uprho^{0}$ meson, obtained by combining $\uppi^{+}\uppi^{-}$ pairs using their rest mass, as reported in the PDG~\cite{PDGpionMass}. The spectrum, parameterized with a Breit--Wigner function, is used for the purity estimation of the $\uprho^{0}$ candidates (green). The K$^{0}_{\mathrm{S}}$ contribution (orange) is modeled with a Gaussian function. The $\omega$(782) contribution (violet) is extracted from simulation. The residual background (blue) is parameterized with an exponential function. The $\uprho^{0}$ candidates are selected within the mass window $0.70$--$0.85$~\GeVmass (indicated by long-dashed lines). The dotted lines indicate the signal and sideband regions, respectively.}
\label{fig:Minv_with_fit}
\end{figure}

The purity of the $\uprho^{0}$ candidate sample is enhanced by requiring a minimum transverse momentum of the $\uprho^{0}$ candidate of $1.80$~\GeVc. The invariant-mass (\Minv) spectrum is obtained by combining $\uppi^{+}\uppi^{-}$ pairs, using their known rest mass. The resulting \Minv distribution is studied differentially as a function of the transverse momentum of the $\uprho^{0}$ candidate. The purity is estimated by modeling \Minv based on the methods in Refs.~\cite{STAR:2003vqj, ALICE:2018qdv}. One example of the \Minv is shown in Fig.~\ref{fig:Minv_with_fit}. For improved visibility, the like-sign background, constructed from the geometric mean of $\uppi^{+}\uppi^{+}$ and $\uppi^{-}\uppi^{-}$ pairs, is subtracted, removing a substantial amount of the combinatorial background. The resonance peaks for the K$^{0}_{\mathrm{S}}$ as well as $\uprho^{0}$ and a hint at the two-body decay channel of the $\omega$(782) are visible. Below $\Minv = 0.45$~\GeVmass the difference of the invariant-masses becomes sensitive to the final-state interaction between the pairs, eventually leading to negative values. As this region is far from the signal, the lower limit for the fit to the invariant mass spectrum is $0.45$~\GeVmass, which is represented by the red solid line. In previous studies~\cite{OPAL:1998enc, DELPHI:1994qgk, BEDDALL2009300} the shape of the blue-coloured background, $f_{\rm LSB}(\Minv)~=~(\Minv - 2m_\uppi)^{n}\exp(\mathcal{A} + \mathcal{B}\Minv + \mathcal{C}\Minv^2)$ was shown to reproduce the residual correlated background, free of resonances, produced in MC studies. The width of the K$^{0}_{\mathrm{S}}$ contribution, represented in orange, is purely dominated by the momentum resolution of the detector, hence a Gaussian shape is assumed. The $\omega$(782) contribution is shown in violet and parametrized using a scaled template, obtained from MC simulations using the identical setup as for the DCA$_{xy}$ templates. The broad structure of this distribution arises from combining the charged pions of the three-body decay channel of $\upomega$ ($\upomega\mathrm{(782)}\rightarrow\uppi^{+}\uppi^{-}\uppi^{0}$). The $\uprho^{0}$ signal, depicted in green, is assumed to follow a relativistic Breit-Wigner distribution~\cite{PhysRev.49.519, PDG2024_reso} with an energy-dependent width. The upper fit range is limited to $0.90$~\GeVmass to reduce the complexity of the fit as the f$_{0}$(980) contribution can be neglected. A crosscheck including the f$_{0}$(980) was performed by enlarging the fit range to $1.10$~\GeVmass, confirming that the upper limit of $0.90$~\GeVmass introduces no significant bias. The model's compatibility with the data is verified by computing the reduced $\chi^2$  metric yielding a value of $4.12$. The $\uprho^{0}$ candidates are selected within the window of $0.70$--$0.85$~\GeVmass (indicated by vertical long-dashed lines) with a \pt averaged purity of $3.26\%$. In the sideband (SB) regions (indicated by vertical short-dashed lines), defined as $0.67$--$0.70$~\GeVmass and $0.85$--$0.88$~\GeVmass, the \pt averaged purity is $1.50\%$.

Since the $N_{\mathrm{same}}(\ks)$ and $N_{\mathrm{mixed}}(\ks)$ for \rhop and \rhoap are consistent within their respective uncertainties, \Cks is computed from the direct sum of the same event samples and in the following \rhop will refer to \rhop$\oplus$\rhoap. The correlation function is normalised to unity in $\ks\in[600, 800]$~\MeVc, where \Cks is expected to be void of any correlations. In practice, these measurements contain all correlations imprinted on the relative momentum (\ks) distribution regardless of the origin. Consequently any background contribution such as feed down, misidentification, and physical background sources needs to be accounted for. Especially, in meson$-$baryon systems, measurements are hampered by correlated residual background which stems from hard partonic interactions in the initial stage of the collision producing highly correlated particles in jet-like structures (often referred to as minijets)~\cite{ALICE:2022yyh}, resulting in events that are generally less isotropic. The experimentally measured \Cks is therefore decomposed as follows
\begin{equation}
C^{\rm{exp}}(k^*) =C^{\mathrm{minijet}}(k^*) \left[\lambda_{\rhop} C_{\rhop}(k^*) + \lambda_{\rm{flat}}C_{\rm{flat}}(k^*)\right] + \lambda_{\widetilde{\uprho}^0\mbox{--}\rm{p}}C_{\widetilde{\uprho}^0\mbox{--}\rm{p}}(k^*),
\label{eq:cf_Sig_general}
\end{equation}
where $C^{\mathrm{minijet}}(k^*)$ denotes the correlation present due to the minijet background, $C_{\rhop}(k^*)$ is the genuine \rhop correlation function, $C_{\rm{flat}}$ entails all contribution due to misidentifcation of the proton and/or $\uprho^{0}$ and is assumed to equal unity. Finally, $C_{\widetilde{\uprho}^0\mbox{--}\rm{p}}(k^*)$ subsumes all contributions for which two oppositely charged pions are paired with a proton. The relative weight of each contribution is encoded in the $\lambda-$parameter~\cite{Lisa:2005dd, 2021ARNPS..71..377F}. Their values are determined in a data-driven way from single-particle properties~\cite{ALICE:2018ysd} (purity and primary fraction) and summarized in Tab.~\ref{tab:lambdaParam}.
\begin{table}[h!]
\begin{center}
\caption{Weight parameters used for the decomposition of the \rhop correlation function.}
\begin{tabular}{l  c  c }
\hline\hline
$\lambda_{i-j}$ & signal region [\%] & sideband [\%]\\
\hline 
\rhop & 2.7 & 1.3\\
flat & 1.5 & 1.5\\
$\widetilde{\uprho}^0\mbox{--}\rm{p} $& 81.4 & 82.9\\
\hline \hline
\end{tabular}
\label{tab:lambdaParam}
\end{center}
\end{table}
An estimate for $C_{\widetilde{\uprho}^0\mbox{--}\rm{p}}$, is conveniently obtained from the measured sidebands $(C^{\rm{exp, SB}}(\ks))$. Assuming the same general decomposition in Eq.~\eqref{eq:cf_Sig_general} yields
\begin{align}
C^{\rm{exp, SB}}(\ks) &= \left[\omega_{\rm{left}}C^{\mathrm{left, SB}}(\ks) + (1-\omega_{\rm{left}})C^{\mathrm{right, SB}}(\ks)\right] \nonumber \\
&= C^{\mathrm{minijet}}(\ks) \left[\lambda_{\rhop}^{\rm{SB}}C_{\rhop}(\ks) + \lambda_{\rm{flat}}^{\rm{SB}}\right] 
+ \lambda_{\widetilde{\uprho}^0\mbox{--}\rm{p}}^{\rm{SB}}C_{\widetilde{\uprho}^0\mbox{--}\rm{p}}(\ks),
\label{eq:cf_SB_general}
\end{align}
where $\omega_{\rm{left}}$ ($=50$\%) represents the relative weight of the left sideband contribution. This weight is determined from the yield of the background underneath the signal peak in the full invariant-mass distribution. Specifically, $\omega_{\rm{left}}$ is obtained by integrating the background over the left and right halves of the signal region, which is symmetrically split around the peak, and computing the relative fraction. The construction of the sideband correlation is verified with a dedicated MC closure study such that the true background correlation is compared with the estimated background shape obtained employing the procedure described above. The closure is achieved within 2.5\% for \ks larger than $500$~\MeVc, which is accounted for as a normalization uncertainty. Now Eq.~\eqref{eq:cf_SB_general} can be solved for $C_{\widetilde{\\\uprho}^0\mbox{--}\rm{p}}$ and the resulting expression is substituted into Eq.~\eqref{eq:cf_Sig_general}. After grouping the terms and solving for the genuine correlation of \rhop carried by $C_{\rhop}(k^*)$, one obtains
\begin{equation}
 \left(\lambda_{\rhop} - \lambda_{\rhop}^{\rm{SB}}\frac{\lambda_{\widetilde{\uprho}^0\mbox{--}\rm{p}}}{\lambda^{\rm{SB}}_{\widetilde{\uprho}^0\mbox{--}\rm{p}}} \right)C_{\rhop}(\ks) = \frac{1}{C^{\mathrm{minijet}}(\ks)} \left[ C^{\rm{exp}}(\ks) - \frac{\lambda_{\widetilde{\uprho}^0\mbox{--}\rm{p}}}{\lambda^{\rm{SB}}_{\widetilde{\uprho}^0\mbox{--}\rm{p}}}  C^{\rm{exp, SB}}(\ks)\right] - \left( 1- \frac{\lambda_{\widetilde{\uprho}^0\mbox{--}\rm{p}}}{\lambda^{\rm{SB}}_{\widetilde{\uprho}^0\mbox{--}\rm{p}}} \right)\lambda_{\rm{flat}}.
\label{eq:cf_rhopGen_pSBcorr_full}
\end{equation}
Inspecting the magnitude of the last term yields that the flat contribution is on the order of sub-promille, and negligible within the experimental uncertainties. Furthermore, the fraction $\lambda_{\widetilde{\uprho}^0\mbox{--}\rm{p}}/\lambda^{\rm{SB}}_{\widetilde{\uprho}^0\mbox{--}\rm{p}}$ simplifies as expected to the ratio of $\uprho^{0}$ impurities, as the selection criteria for the proton do not change for the measurement in the signal and sideband region. The prefactor in front of $C_{\rhop}(k^*)$ carries no physical relevance, because the normalization of the correlation function is usually included as a free parameter in model comparisons. Lastly, $C^{\mathrm{minijet}}(\ks)$ is modeled using the averaged sideband correlation $C^{\rm{exp, SB}}(\ks)$ assuming that $C^{\rm{exp, SB}}(\ks)$ is dominated by the minijet contribution. The latter assumption was verified by projecting the theoretical pair-wise two-body correlations in the triplet $(\pi^{+}\pi^{-}\rm{p})$ into the kinematic frame of the \rhop, using the technique introduced in Ref.~\cite{DelGrande:2021mju}. The resulting correlation function shows a negligible deviation from unity, as expected in the absence of final-state interactions. This demonstrates that the measured $C^{\rm{exp, SB}}(\ks)$ is dominated by the minijet contribution. 

The systematic uncertainties for the obtained correlation function are estimated by varying the selection criteria (including standard track quality selections) for the proton and charged pions, respectively, and the \Minv ranges for the $\uprho^{0}$ candidates, such that any combination does not exceed a variation of $20$\% of the pair yield in $N_{\mathrm{same}}(\ks)$ for \ks below $200$~\MeVc. The latter condition ensures statistical significance. For each $\ks$ range, the systematic uncertainty is then estimated as the standard deviation of the resulting distribution of correlation values across all accepted variations. The underlying distribution is assumed to be uniform.

\begin{figure}[hbt]
\centering
\includegraphics[width=.5\textwidth]{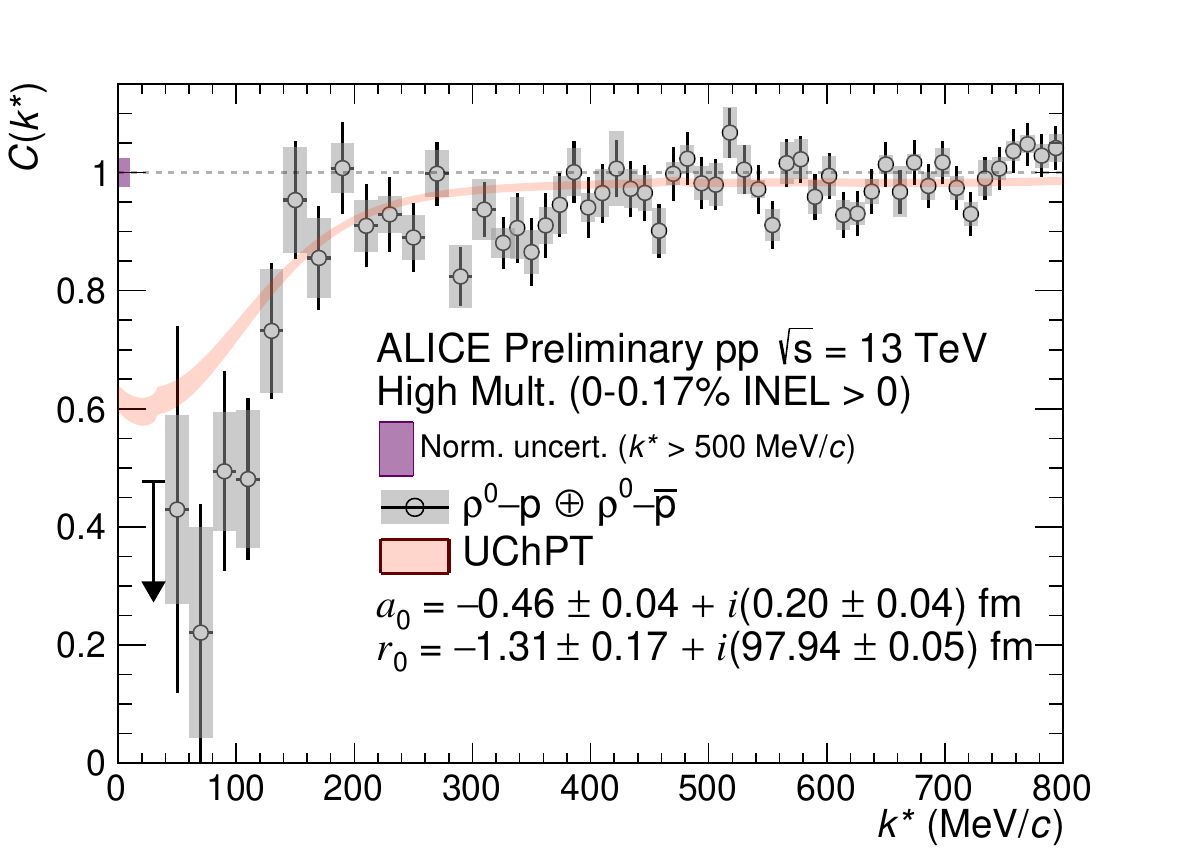}
\caption{Genuine \rhop correlation function. Statistical uncertainties are shown as error bars, and systematic uncertainties as shaded rectangles. The uncertainty band of the fit, based on unitarized chiral perturbation theory~\cite{Feijoo:2024bvn}, is determined using a bootstrap procedure.}
\label{fig:CF_with_fit}
\end{figure}

The genuine correlation function, unfolded for purity and background contributions according to Eq.~\eqref{eq:cf_rhopGen_pSBcorr_full}, is presented in Fig.~\ref{fig:CF_with_fit}. For \ks above $200$~\MeVc, the correlation function remains consistent with unity within uncertainties, reflecting the expected absence of strong interaction effects at large relative momentum. Below $200$~\MeVc, where the FSI effects are expected, the data exhibit a $4\upsigma$ suppression relative to unity, providing a clear and strong signature of the $\uprho^0$p interaction

The interpretation of the correlation function needs to account for coupled channel contributions to encapsulate the full interaction dynamics, which requires the use of an adapted Koonin-Pratt equation given by $C_i(k^*) = \sum_{j} \omega_{j} ^{\rm prod.} \int d^3 r^* S_{j} (r^*) |\uppsi_{ji} (k^*,r^*)|^2$~\cite{HAIDENBAUER20191, Feijoo:2024bvn}. The correlation function of the measured channel $i$ is obtained by the summation of correlation functions for the $j$ coupled channels ($j= \uprho^0 \rm{p},~ \uprho^+ \rm{n},~ \upomega \rm{p},~ \upphi \rm{p},~ \rm{K}^{*+}\Lambda,~ \rm{K}^{*0}\Sigma^+,~ \rm{K}^{*+}\Sigma^0$). Analogous to the case without coupled channels each channel $j$ requires knowledge of the specific source $S_{j} (r^*)$ and wave function $\uppsi_{ji} (k^*,r^*)$, which satisfies the boundary condition that it evolves into an outgoing wave in channel $i$ at large distances. The contributions are scaled by the production weights $\omega_j^{\rm prod.}$, and evaluated with the data-driven method introduced in Ref.~\cite{ALICE:2022yyh} (more information is available in the appendix~\ref{appendix:ProductionWeights}). An advantage of this ansatz is that the recent \phiP correlation function measurement~\cite{ALICE:2021cpv} can be exploited as an additional constraint for the interpretation of the $\uprho^0$p interaction.

The source function for each channel is obtained by building on the findings of Refs.~\cite{ALICE:2020ibs, ALICE:2023sjd, ALICE:2024Marcel}, which studied $S(r^*)$ for various hadron--hadron pairs in small systems. These works demonstrate that a common emission source can be assumed across different hadron species. This holds, provided feed down from strongly decaying resonances is properly accounted for. At the same time, they provide the detailed dependence of the primordial source, parametrized with a Gaussian width (\rc), on the mean transverse mass (\smash{$\mt = ( k_{\rm{T}}^{2}  + m_{\rm{avrg.}}^{2})^{1/2}$}) of the pair, where the average mass and \pt is denoted by $ m_{\rm{avrg.}}$ and $ k_{\rm{T}}  = \frac{1}{2}\left( \Vec{\pt}_{1} + \Vec{\pt}_{2} \right)$, respectively. The core radius for \rhop and \phiP, as well as for the respective coupled channels, are evaluated by the mean $\mt$ of 2.33~\GeVmass and 1.66~\GeVmass, resulting in \rc of $0.78\pm0.06$~fm and $0.90\pm0.04$~fm, respectively. The strongly decaying resonances are explicitly considered for all particles according to the resonance source model introduced in Ref.~\cite{ALICE:2020ibs}, with additional details provided in the appendix~\ref{appendix:Source}.

The interaction is modeled using calculations from unitarized chiral perturbation theory, incorporating vector-meson interactions via the hidden gauge symmetry formalism within a coupled channel framework, as introduced in Ref.~\cite{Feijoo:2024bvn}. The wave functions are determined from the solution of the Bethe-Salpeter equation~\cite{PhysRev.84.1232} with coupled channels, which yields the full scattering matrix $T$. The correlation functions are evaluated according to the modified version of the Koonin-Pratt equation. The model has 6 free parameters: 5 subtractions constants (SCs) and a normalization (see appendix~\ref{appendix:SubtractionConstants}). The values for the SCs are obtained by minimizing simultaneously the reduced $\chi^2$ metrics of the models to the measured $C_{\rhop}(\ks)$ and $C_{\phiP}(\ks)$, using the corresponding $\omega_{j}^{\rm prod.}$ and $S_{j}(\rs)$. The full uncertainties are obtained by employing the bootstrap technique~\cite{Efron:1986hys}, following the implementation in Ref.~\cite{ALICE:2023sjd}. The scattering parameters of \rhop are extracted by analyzing the s-wave phase-shift $\delta_0(\ks)$ of the obtained wave function. At low energies $\delta_0(\ks)$ can be expanded in powers of \ks, which leads to the well-known relation 
$\lim_{\ks \to 0} \ks \cot{\delta_0(\ks)} = -(a_{0})^{-1}+ \frac{1}{2}r_{0}\ksSq$,
encapsulating the information about the interaction in the two scattering parameters: the scattering length $a_{0}$ and the effective range $r_{0}$. This analysis yields a scattering length of $a_{\uprho^0\mathrm{p}} = (-0.46 \pm 0.04) + i(0.20 \pm 0.04)$~fm and an effective range of $r_{\uprho^0\mathrm{p}} = (-1.31 \pm 0.17) + i(97.94 \pm 0.05)$~fm. The width of the $\uprho^{0}$ is encoded in the imaginary part of $r_{\uprho^0\mathrm{p}}$, naturally leading to a notably large value. This finding is consistent with a QCD sum rule calculation~\cite{Koike:1996ga} $a_{\uprho^0\mathrm{p}} = (-0.40 \pm 0.05)$~fm and the modulus is in agreement with a recent analysis of photoproduction data using VMD~\cite{Wang:2022zwz} $\lvert{a_{\uprho^0\mathrm{p}}}\rvert = (0.36 \pm 0.06)$~fm (since VMDs are restricted to the modulus). Deviations in the scattering lengths highlight the necessity of direct two-body \rhop measurements, offering a stringent constraint on the vacuum interaction dynamics.

Analogous to the procedure in Ref.~\cite{Feijoo:2024bvn}, two poles, identified with the N(1958) and N(1700) resonances, are found in the second Riemann sheet of $T$. Their positions align with previous findings~\cite{Feijoo:2024bvn, Oset:2010tof, Khemchandani:2011et, Gamermann:2011mq, Garzon:2012np}, supporting the robustness of the extracted scattering parameters. While a full decomposition into coupling strengths is beyond the scope of this work, this verification ensures that the obtained scattering parameters correctly reflect the underlying interaction dynamics. Traditionally, spectroscopic information of this kind is obtained by interpreting data from low-energy experiments, such as those conducted by HADES, within a partial wave analysis framework~\cite{PhysRevC.106.064910, HADES:2020kce}. However, these approaches are limited in accessing states below the production threshold, underscoring the complementary strength of the presented methodology. 

This work introduces a complementary approach to studying resonance states at low energies and provides crucial inputs for refining vector meson--baryon interaction models. The reported results will serve as valuable input for pertinent in-medium calculation frameworks. The upcoming Run~3 and Run~4 data taking at the LHC will significantly improve the precision of the extracted interaction parameters. This measurement demonstrates, for the first time, how to directly access the \rhop interaction and offers essential experimental insight into the $\uprho$-meson–nucleon interaction in vacuum. This marks a cornerstone for achieving a self-consistent description of the $\uprho$-meson’s in-medium properties, which is crucial for interpreting heavy-ion collision data.


\newenvironment{acknowledgement}{\relax}{\relax}
\begin{acknowledgement}
\section*{Acknowledgements}
The ALICE Collaboration is very grateful to A. Feijoo for the valuable discussions and guidance on meson-baryon interactions.

The ALICE Collaboration would like to thank all its engineers and technicians for their invaluable contributions to the construction of the experiment and the CERN accelerator teams for the outstanding performance of the LHC complex.
The ALICE Collaboration gratefully acknowledges the resources and support provided by all Grid centres and the Worldwide LHC Computing Grid (WLCG) collaboration.
The ALICE Collaboration acknowledges the following funding agencies for their support in building and running the ALICE detector:
A. I. Alikhanyan National Science Laboratory (Yerevan Physics Institute) Foundation (ANSL), State Committee of Science and World Federation of Scientists (WFS), Armenia;
Austrian Academy of Sciences, Austrian Science Fund (FWF): [M 2467-N36] and Nationalstiftung f\"{u}r Forschung, Technologie und Entwicklung, Austria;
Ministry of Communications and High Technologies, National Nuclear Research Center, Azerbaijan;
Rede Nacional de Física de Altas Energias (Renafae), Financiadora de Estudos e Projetos (Finep), Funda\c{c}\~{a}o de Amparo \`{a} Pesquisa do Estado de S\~{a}o Paulo (FAPESP) and The Sao Paulo Research Foundation  (FAPESP), Brazil;
Bulgarian Ministry of Education and Science, within the National Roadmap for Research Infrastructures 2020-2027 (object CERN), Bulgaria;
Ministry of Education of China (MOEC) , Ministry of Science \& Technology of China (MSTC) and National Natural Science Foundation of China (NSFC), China;
Ministry of Science and Education and Croatian Science Foundation, Croatia;
Centro de Aplicaciones Tecnol\'{o}gicas y Desarrollo Nuclear (CEADEN), Cubaenerg\'{\i}a, Cuba;
Ministry of Education, Youth and Sports of the Czech Republic, Czech Republic;
The Danish Council for Independent Research | Natural Sciences, the VILLUM FONDEN and Danish National Research Foundation (DNRF), Denmark;
Helsinki Institute of Physics (HIP), Finland;
Commissariat \`{a} l'Energie Atomique (CEA) and Institut National de Physique Nucl\'{e}aire et de Physique des Particules (IN2P3) and Centre National de la Recherche Scientifique (CNRS), France;
Bundesministerium f\"{u}r Bildung und Forschung (BMBF) and GSI Helmholtzzentrum f\"{u}r Schwerionenforschung GmbH, Germany;
General Secretariat for Research and Technology, Ministry of Education, Research and Religions, Greece;
National Research, Development and Innovation Office, Hungary;
Department of Atomic Energy Government of India (DAE), Department of Science and Technology, Government of India (DST), University Grants Commission, Government of India (UGC) and Council of Scientific and Industrial Research (CSIR), India;
National Research and Innovation Agency - BRIN, Indonesia;
Istituto Nazionale di Fisica Nucleare (INFN), Italy;
Japanese Ministry of Education, Culture, Sports, Science and Technology (MEXT) and Japan Society for the Promotion of Science (JSPS) KAKENHI, Japan;
Consejo Nacional de Ciencia (CONACYT) y Tecnolog\'{i}a, through Fondo de Cooperaci\'{o}n Internacional en Ciencia y Tecnolog\'{i}a (FONCICYT) and Direcci\'{o}n General de Asuntos del Personal Academico (DGAPA), Mexico;
Nederlandse Organisatie voor Wetenschappelijk Onderzoek (NWO), Netherlands;
The Research Council of Norway, Norway;
Pontificia Universidad Cat\'{o}lica del Per\'{u}, Peru;
Ministry of Science and Higher Education, National Science Centre and WUT ID-UB, Poland;
Korea Institute of Science and Technology Information and National Research Foundation of Korea (NRF), Republic of Korea;
Ministry of Education and Scientific Research, Institute of Atomic Physics, Ministry of Research and Innovation and Institute of Atomic Physics and Universitatea Nationala de Stiinta si Tehnologie Politehnica Bucuresti, Romania;
Ministerstvo skolstva, vyskumu, vyvoja a mladeze SR, Slovakia;
National Research Foundation of South Africa, South Africa;
Swedish Research Council (VR) and Knut \& Alice Wallenberg Foundation (KAW), Sweden;
European Organization for Nuclear Research, Switzerland;
Suranaree University of Technology (SUT), National Science and Technology Development Agency (NSTDA) and National Science, Research and Innovation Fund (NSRF via PMU-B B05F650021), Thailand;
Turkish Energy, Nuclear and Mineral Research Agency (TENMAK), Turkey;
National Academy of  Sciences of Ukraine, Ukraine;
Science and Technology Facilities Council (STFC), United Kingdom;
National Science Foundation of the United States of America (NSF) and United States Department of Energy, Office of Nuclear Physics (DOE NP), United States of America.
In addition, individual groups or members have received support from:
Czech Science Foundation (grant no. 23-07499S), Czech Republic;
FORTE project, reg.\ no.\ CZ.02.01.01/00/22\_008/0004632, Czech Republic, co-funded by the European Union, Czech Republic;
European Research Council (grant no. 950692), European Union;
Deutsche Forschungs Gemeinschaft (DFG, German Research Foundation) ``Neutrinos and Dark Matter in Astro- and Particle Physics'' (grant no. SFB 1258), Germany;
ICSC - National Research Center for High Performance Computing, Big Data and Quantum Computing and FAIR - Future Artificial Intelligence Research, funded by the NextGenerationEU program (Italy).

\end{acknowledgement}

\bibliographystyle{utphys}   
\bibliography{bibliography}

\providecommand{\href}[2]{#2}\begingroup\raggedright\begin{thebibliography}{10}

\bibitem{PhysRevC.46.R34}
T.~Hatsuda and S.~H. Lee, ``{QCD} sum rules for vector mesons in the nuclear
  medium'', \href{https://doi.org/10.1103/PhysRevC.46.R34}{{\em Phys. Rev. C}
  {\bfseries 46} (Jul, 1992) R34--R38}.
  \url{https://link.aps.org/doi/10.1103/PhysRevC.46.R34}.

\bibitem{doi:10.1142/S0218301310014728}
S.~Leupold, V.~Metag, and U.~Mosel, ``Hadrons in strongly interacting matter'',
  \href{https://doi.org/10.1142/S0218301310014728}{{\em International Journal
  of Modern Physics E} {\bfseries 19} (2010) 147--224}.
  \url{https://doi.org/10.1142/S0218301310014728}.

\bibitem{Zschocke2002}
S.~Zschocke, O.~Pavlenko, and B.~Kämpfer, ``Evaluation of {QCD} sum rules for
  light vector mesons at finite density and temperature'',
  \href{https://doi.org/10.1140/epja/i2002-10062-4}{{\em European Physical
  Journal A} {\bfseries 15} (12, 2002) 529--537}.

\bibitem{PhysRevLett.66.2720}
G.~E. Brown and M.~Rho, ``Scaling effective lagrangians in a dense medium'',
  \href{https://doi.org/10.1103/PhysRevLett.66.2720}{{\em Phys. Rev. Lett.}
  {\bfseries 66} (May, 1991) 2720--2723}.
  \url{https://link.aps.org/doi/10.1103/PhysRevLett.66.2720}.

\bibitem{BROCKMANN199640}
R.~Brockmann and W.~Weise, ``The chiral condensate in nuclear matter'',
  \href{https://doi.org/https://doi.org/10.1016/0370-2693(95)01448-9}{{\em
  Physics Letters B} {\bfseries 367} (1996) 40--44}.
  \url{https://www.sciencedirect.com/science/article/pii/0370269395014489}.

\bibitem{Rapp2000}
R.~Rapp and J.~Wambach, {\em Chiral Symmetry Restoration and Dileptons in
  Relativistic Heavy-Ion Collisions},
  \href{https://doi.org/10.1007/0-306-47101-9_1}{pp.~1--205}.
\newblock Springer US, Boston, MA, 2000.
\newblock \url{https://doi.org/10.1007/0-306-47101-9_1}.

\bibitem{refId0}
V.~Metag, M.~Nanova, and K.-T. Brinkmann, ``In-medium properties of mesons'',
  \href{https://doi.org/10.1051/epjconf/201713403003}{{\em EPJ Web Conf.}
  {\bfseries 134} (2017) 03003}.
  \url{https://doi.org/10.1051/epjconf/201713403003}.

\bibitem{ALICE:2022wpn}
{\bfseries ALICE} Collaboration, S.~Acharya {\em et~al.}, ``{The ALICE
  experiment: a journey through QCD}'',
  \href{https://doi.org/10.1140/epjc/s10052-024-12935-y}{{\em Eur. Phys. J. C}
  {\bfseries 84} (2024) 813},
  \href{https://arxiv.org/abs/2211.04384}{{\ttfamily arXiv:2211.04384
  [nucl-ex]}}.

\bibitem{Dusling:2006yv}
K.~Dusling, D.~Teaney, and I.~Zahed, ``{Thermal dimuon yields at NA60}'',
  \href{https://doi.org/10.1103/PhysRevC.75.024908}{{\em Phys. Rev. C}
  {\bfseries 75} (2007) 024908},
  \href{https://arxiv.org/abs/nucl-th/0604071}{{\ttfamily
  arXiv:nucl-th/0604071}}.

\bibitem{Ruppert:2006hf}
J.~Ruppert and T.~Renk, ``{What does the rho-meson do? In-medium mass shift
  scenarios versus hadronic model calculations}'',
  \href{https://doi.org/10.1140/epjc/s10052-006-0072-y}{{\em Eur. Phys. J. C}
  {\bfseries 49} (2007) 219--224},
  \href{https://arxiv.org/abs/nucl-th/0609083}{{\ttfamily
  arXiv:nucl-th/0609083}}.

\bibitem{CERES:2006wcq}
{\bfseries CERES} Collaboration, D.~Adamova {\em et~al.}, ``{Modification of
  the rho-meson detected by low-mass electron-positron pairs in central Pb-Au
  collisions at 158$A$GeV/$c$}'',
  \href{https://doi.org/10.1016/j.physletb.2008.07.104}{{\em Phys. Lett. B}
  {\bfseries 666} (2008) 425--429},
  \href{https://arxiv.org/abs/nucl-ex/0611022}{{\ttfamily
  arXiv:nucl-ex/0611022}}.

\bibitem{NA60:2006ymb}
{\bfseries NA60} Collaboration, R.~Arnaldi {\em et~al.}, ``{First measurement
  of the rho spectral function in high-energy nuclear collisions}'',
  \href{https://doi.org/10.1103/PhysRevLett.96.162302}{{\em Phys. Rev. Lett.}
  {\bfseries 96} (2006) 162302},
  \href{https://arxiv.org/abs/nucl-ex/0605007}{{\ttfamily
  arXiv:nucl-ex/0605007}}.

\bibitem{HADES:2020kce}
{\bfseries HADES} Collaboration, J.~Adamczewski-Musch {\em et~al.}, ``{Two-pion
  production in the second resonance region in ${\pi}^-p$ collisions with the
  High-Acceptance Di-Electron Spectrometer (HADES)}'',
  \href{https://doi.org/10.1103/PhysRevC.102.024001}{{\em Phys. Rev. C}
  {\bfseries 102} (2020) 024001},
  \href{https://arxiv.org/abs/2004.08265}{{\ttfamily arXiv:2004.08265
  [nucl-ex]}}.

\bibitem{FRIMAN1997496}
B.~Friman and H.~Pirner, ``P-wave polarization of the $\rho$-meson and the
  dilepton spectrum in dense matter'',
  \href{https://doi.org/https://doi.org/10.1016/S0375-9474(97)00050-X}{{\em
  Nuclear Physics A} {\bfseries 617} (1997) 496--509}.
  \url{https://www.sciencedirect.com/science/article/pii/S037594749700050X}.

\bibitem{PHENIX:2008qav}
{\bfseries PHENIX} Collaboration, A.~Adare {\em et~al.}, ``{Dilepton mass
  spectra in p+p collisions at s**(1/2) = 200-GeV and the contribution from
  open charm}'', \href{https://doi.org/10.1016/j.physletb.2008.10.064}{{\em
  Phys. Lett. B} {\bfseries 670} (2009) 313--320},
  \href{https://arxiv.org/abs/0802.0050}{{\ttfamily arXiv:0802.0050 [hep-ex]}}.

\bibitem{PHENIX:2009gyd}
{\bfseries PHENIX} Collaboration, A.~Adare {\em et~al.}, ``{Detailed
  measurement of the $e^+ e^-$ pair continuum in $p+p$ and Au+Au collisions at
  $\sqrt{s_{NN}} = 200$ GeV and implications for direct photon production}'',
  \href{https://doi.org/10.1103/PhysRevC.81.034911}{{\em Phys. Rev. C}
  {\bfseries 81} (2010) 034911},
  \href{https://arxiv.org/abs/0912.0244}{{\ttfamily arXiv:0912.0244
  [nucl-ex]}}.

\bibitem{PHENIX:2015vek}
{\bfseries PHENIX} Collaboration, A.~Adare {\em et~al.}, ``{Dielectron
  production in Au$+$Au collisions at $\sqrt{s_{NN}}$=200 GeV}'',
  \href{https://doi.org/10.1103/PhysRevC.93.014904}{{\em Phys. Rev. C}
  {\bfseries 93} (2016) 014904},
  \href{https://arxiv.org/abs/1509.04667}{{\ttfamily arXiv:1509.04667
  [nucl-ex]}}.

\bibitem{STAR:2012dzw}
{\bfseries STAR} Collaboration, L.~Adamczyk {\em et~al.}, ``{Di-electron
  spectrum at mid-rapidity in $p+p$ collisions at $\sqrt{s} = 200$ GeV}'',
  \href{https://doi.org/10.1103/PhysRevC.86.024906}{{\em Phys. Rev. C}
  {\bfseries 86} (2012) 024906},
  \href{https://arxiv.org/abs/1204.1890}{{\ttfamily arXiv:1204.1890
  [nucl-ex]}}.

\bibitem{STAR:2013pwb}
{\bfseries STAR} Collaboration, L.~Adamczyk {\em et~al.}, ``{Dielectron Mass
  Spectra from Au+Au Collisions at $\sqrt{s_{\rm NN}}$ = 200 GeV}'',
  \href{https://doi.org/10.1103/PhysRevLett.113.022301}{{\em Phys. Rev. Lett.}
  {\bfseries 113} (2014) 022301},
  \href{https://arxiv.org/abs/1312.7397}{{\ttfamily arXiv:1312.7397 [hep-ex]}}.
  [Addendum: Phys.Rev.Lett. 113, 049903 (2014)].

\bibitem{Huck:2014mfa}
{\bfseries STAR} Collaboration, P.~Huck, ``{Beam Energy Dependence of
  Dielectron Production in Au $+$ Au Collisions from STAR at RHIC}'',
  \href{https://doi.org/10.1016/j.nuclphysa.2014.09.090}{{\em Nucl. Phys. A}
  {\bfseries 931} (2014) 659--664},
  \href{https://arxiv.org/abs/1409.5675}{{\ttfamily arXiv:1409.5675
  [nucl-ex]}}.

\bibitem{STAR:2015tnn}
{\bfseries STAR} Collaboration, L.~Adamczyk {\em et~al.}, ``{Measurements of
  Dielectron Production in Au$+$Au Collisions at $\sqrt{s_{\rm NN}}$ = 200 GeV
  from the STAR Experiment}'',
  \href{https://doi.org/10.1103/PhysRevC.92.024912}{{\em Phys. Rev. C}
  {\bfseries 92} (2015) 024912},
  \href{https://arxiv.org/abs/1504.01317}{{\ttfamily arXiv:1504.01317
  [hep-ex]}}.

\bibitem{STAR:2015zal}
{\bfseries STAR} Collaboration, L.~Adamczyk {\em et~al.}, ``{Energy dependence
  of acceptance-corrected dielectron excess mass spectrum at mid-rapidity in
  Au$+$Au collisions at $\sqrt{s_{NN}} =$ 19.6 and 200 GeV}'',
  \href{https://doi.org/10.1016/j.physletb.2015.08.044}{{\em Phys. Lett. B}
  {\bfseries 750} (2015) 64--71},
  \href{https://arxiv.org/abs/1501.05341}{{\ttfamily arXiv:1501.05341
  [hep-ex]}}.

\bibitem{STAR:2018ldd}
{\bfseries STAR} Collaboration, J.~Adam {\em et~al.}, ``{Low-$p_T$ $e^{+}e^{-}$
  pair production in Au$+$Au collisions at $\sqrt{s_{NN}}$ = 200 GeV and U$+$U
  collisions at $\sqrt{s_{NN}}$ = 193 GeV at STAR}'',
  \href{https://doi.org/10.1103/PhysRevLett.121.132301}{{\em Phys. Rev. Lett.}
  {\bfseries 121} (2018) 132301},
  \href{https://arxiv.org/abs/1806.02295}{{\ttfamily arXiv:1806.02295
  [hep-ex]}}.

\bibitem{STAR:2023wta}
{\bfseries STAR} Collaboration, M.~I. Abdulhamid {\em et~al.}, ``{Measurements
  of dielectron production in Au+Au collisions at sNN=27, 39, and 62.4 GeV from
  the STAR experiment}'',
  \href{https://doi.org/10.1103/PhysRevC.107.L061901}{{\em Phys. Rev. C}
  {\bfseries 107} (2023) L061901}.

\bibitem{ALICE:2018ael}
{\bfseries ALICE} Collaboration, S.~Acharya {\em et~al.}, ``{Measurement of
  dielectron production in central Pb--Pb collisions at
  $\sqrt{{\textit{s}}_{\mathrm{NN}}}$ = 2.76 TeV}'',
  \href{https://doi.org/10.1103/PhysRevC.99.024002}{{\em Phys. Rev. C}
  {\bfseries 99} (2019) 024002},
  \href{https://arxiv.org/abs/1807.00923}{{\ttfamily 1807.00923 [nucl-ex]}}.

\bibitem{ALICE:2023jef}
{\bfseries ALICE} Collaboration, S.~Acharya {\em et~al.}, ``{Dielectron
  production in central Pb$-$Pb collisions at $\sqrt{s_\mathrm{NN}}$ = 5.02
  TeV}'', \href{https://arxiv.org/abs/2308.16704}{{\ttfamily arXiv:2308.16704
  [nucl-ex]}}.

\bibitem{NA60:2008ctj}
{\bfseries NA60} Collaboration, R.~Arnaldi {\em et~al.}, ``{NA60 results on
  thermal dimuons}'',
  \href{https://doi.org/10.1140/epjc/s10052-009-0878-5}{{\em Eur. Phys. J. C}
  {\bfseries 61} (2009) 711--720},
  \href{https://arxiv.org/abs/0812.3053}{{\ttfamily arXiv:0812.3053
  [nucl-ex]}}.

\bibitem{Rapp:1997ei}
R.~Rapp, M.~Urban, M.~Buballa, and J.~Wambach, ``{A Microscopic calculation of
  photoabsorption cross-sections on protons and nuclei}'',
  \href{https://doi.org/10.1016/S0370-2693(97)01360-9}{{\em Phys. Lett. B}
  {\bfseries 417} (1998) 1--6},
  \href{https://arxiv.org/abs/nucl-th/9709008}{{\ttfamily
  arXiv:nucl-th/9709008}}.

\bibitem{Rapp:1999us}
R.~Rapp and J.~Wambach, ``{Low mass dileptons at the CERN SPS: Evidence for
  chiral restoration?}'', \href{https://doi.org/10.1007/s100500050364}{{\em
  Eur. Phys. J. A} {\bfseries 6} (1999) 415--420},
  \href{https://arxiv.org/abs/hep-ph/9907502}{{\ttfamily
  arXiv:hep-ph/9907502}}.

\bibitem{PhysRevLett.22.981}
J.~J. Sakurai, ``Vector-meson dominance and high-energy electron-proton
  inelastic scattering'',
  \href{https://doi.org/10.1103/PhysRevLett.22.981}{{\em Phys. Rev. Lett.}
  {\bfseries 22} (May, 1969) 981--984}.
  \url{https://link.aps.org/doi/10.1103/PhysRevLett.22.981}.

\bibitem{Schildknecht:2005xr}
D.~Schildknecht, ``{Vector meson dominance}'', {\em Acta Phys. Polon. B}
  {\bfseries 37} (2006) 595--608,
  \href{https://arxiv.org/abs/hep-ph/0511090}{{\ttfamily
  arXiv:hep-ph/0511090}}.

\bibitem{Rapp:2013nxa}
R.~Rapp, ``{Dilepton Spectroscopy of QCD Matter at Collider Energies}'',
  \href{https://doi.org/10.1155/2013/148253}{{\em Adv. High Energy Phys.}
  {\bfseries 2013} (2013) 148253},
  \href{https://arxiv.org/abs/1304.2309}{{\ttfamily arXiv:1304.2309 [hep-ph]}}.

\bibitem{Hohler:2013eba}
P.~M. Hohler and R.~Rapp, ``{Is $\rho$-Meson Melting Compatible with Chiral
  Restoration?}'', \href{https://doi.org/10.1016/j.physletb.2014.02.021}{{\em
  Phys. Lett. B} {\bfseries 731} (2014) 103--109},
  \href{https://arxiv.org/abs/1311.2921}{{\ttfamily arXiv:1311.2921 [hep-ph]}}.

\bibitem{PhysRevC.106.064910}
K.~Gallmeister, U.~Mosel, and L.~von Smekal, ``Dilepton decay of low-mass
  $\ensuremath{\rho}$ mesons'',
  \href{https://doi.org/10.1103/PhysRevC.106.064910}{{\em Phys. Rev. C}
  {\bfseries 106} (Dec, 2022) 064910}.
  \url{https://link.aps.org/doi/10.1103/PhysRevC.106.064910}.

\bibitem{HADES:2022vus}
{\bfseries HADES} Collaboration, R.~Abou~Yassine {\em et~al.}, ``{First
  measurement of massive virtual photon emission from N* baryon resonances}'',
  \href{https://arxiv.org/abs/2205.15914}{{\ttfamily arXiv:2205.15914
  [nucl-ex]}}.

\bibitem{Feijoo:2024bvn}
A.~Feijoo, M.~Korwieser, and L.~Fabbietti, ``{Relevance of the coupled channels
  in the {\ensuremath{\phi}}p and {\ensuremath{\rho^{0}}}p correlation
  functions}'', \href{https://doi.org/10.1103/PhysRevD.111.014009}{{\em Phys.
  Rev. D} {\bfseries 111} (2025) 014009},
  \href{https://arxiv.org/abs/2407.01128}{{\ttfamily arXiv:2407.01128
  [hep-ph]}}.

\bibitem{ALICE:2020ibs}
{\bfseries ALICE} Collaboration, S.~Acharya {\em et~al.}, ``Corrigendum: Search
  for a common baryon source in high-multiplicity pp collisions at the {LHC}
  (phys. lett. b 811 (2020) 135849)'',
  \href{https://doi.org/https://doi.org/10.1016/j.physletb.2024.139233}{{\em
  Physics Letters B} (2025) 139233}.
  \url{https://www.sciencedirect.com/science/article/pii/S0370269324007913}.

\bibitem{ALICE:2023sjd}
{\bfseries ALICE} Collaboration, S.~Acharya {\em et~al.}, ``{Common femtoscopic
  hadron-emission source in pp collisions at the LHC}'',
  \href{https://doi.org/10.1140/epjc/s10052-025-13793-y}{{\em Eur. Phys. J. C}
  {\bfseries 85} (2025) 198},
  \href{https://arxiv.org/abs/2311.14527}{{\ttfamily arXiv:2311.14527
  [hep-ph]}}.

\bibitem{ALICE:2024Marcel}
{\bfseries ALICE} Collaboration, S.~Acharya {\em et~al.}, ``{Investigating the
  \( p-\pi^{\pm} \) and \( p-p-\pi^{\pm} \) dynamics with femtoscopy in pp
  collisions at \( \sqrt{s} = 13 \) TeV}'', {\em Eur. Phys. J. A} (04, 2025)
  Accepted, \href{https://arxiv.org/abs/2502.20200}{{\ttfamily arXiv:2502.20200
  [hep-ph]}}.

\bibitem{Mihaylov:2023pyl}
D.~Mihaylov and J.~Gonz\'alez~Gonz\'alez, ``{Novel model for particle emission
  in small collision systems}'',
  \href{https://doi.org/10.1140/epjc/s10052-023-11774-7}{{\em Eur. Phys. J. C}
  {\bfseries 83} (2023) 590},
  \href{https://arxiv.org/abs/2305.08441}{{\ttfamily arXiv:2305.08441
  [hep-ph]}}.

\bibitem{ALICE:2020mfd}
{\bfseries ALICE} Collaboration, S.~Acharya {\em et~al.}, ``{Unveiling the
  strong interaction among hadrons at the LHC}'',
  \href{https://doi.org/10.1038/s41586-020-3001-6}{{\em Nature} {\bfseries 588}
  (2020) 232--238}, \href{https://arxiv.org/abs/2005.11495}{{\ttfamily
  arXiv:2005.11495 [nucl-ex]}}. [Erratum: Nature 590, E13 (2021)].

\bibitem{ALICE:2019hdt}
{\bfseries ALICE} Collaboration, S.~Acharya {\em et~al.}, ``{First Observation
  of an Attractive Interaction between a Proton and a Cascade Baryon}'',
  \href{https://doi.org/10.1103/PhysRevLett.123.112002}{{\em Phys. Rev. Lett.}
  {\bfseries 123} (2019) 112002},
  \href{https://arxiv.org/abs/1904.12198}{{\ttfamily arXiv:1904.12198
  [nucl-ex]}}.

\bibitem{ALICE:2022yyh}
{\bfseries ALICE} Collaboration, S.~Acharya {\em et~al.}, ``{Constraining the
  ${\overline{\textrm{K}}}{\textrm{N}}$ coupled channel dynamics using
  femtoscopic correlations at the LHC}'',
  \href{https://doi.org/10.1140/epjc/s10052-023-11476-0}{{\em Eur. Phys. J. C}
  {\bfseries 83} (2023) 340},
  \href{https://arxiv.org/abs/2205.15176}{{\ttfamily arXiv:2205.15176
  [nucl-ex]}}.

\bibitem{ALICE:2021cpv}
{\bfseries ALICE} Collaboration, S.~Acharya {\em et~al.}, ``{Experimental
  Evidence for an Attractive p--$\phi$ Interaction}'',
  \href{https://doi.org/10.1103/PhysRevLett.127.172301}{{\em Phys. Rev. Lett.}
  {\bfseries 127} (2021) 172301},
  \href{https://arxiv.org/abs/2105.05578}{{\ttfamily arXiv:2105.05578
  [nucl-ex]}}.

\bibitem{ALICE:2022enj}
{\bfseries ALICE} Collaboration, S.~Acharya {\em et~al.}, ``{First study of the
  two-body scattering involving charm hadrons}'',
  \href{https://doi.org/10.1103/PhysRevD.106.052010}{{\em Phys. Rev. D}
  {\bfseries 106} (2022) 052010},
  \href{https://arxiv.org/abs/2201.05352}{{\ttfamily arXiv:2201.05352
  [nucl-ex]}}.

\bibitem{ALICE:2024bhk}
{\bfseries ALICE} Collaboration, S.~Acharya {\em et~al.}, ``{Studying the
  interaction between charm and light-flavor mesons}'',
  \href{https://doi.org/10.1103/PhysRevD.110.032004}{{\em Phys. Rev. D}
  {\bfseries 110} (2024) 032004},
  \href{https://arxiv.org/abs/2401.13541}{{\ttfamily arXiv:2401.13541
  [nucl-ex]}}.

\bibitem{Lisa:2005dd}
M.~A. Lisa, S.~Pratt, R.~Soltz, and U.~Wiedemann, ``{Femtoscopy in relativistic
  heavy ion collisions}'',
  \href{https://doi.org/10.1146/annurev.nucl.55.090704.151533}{{\em Ann. Rev.
  Nucl. Part. Sci.} {\bfseries 55} (2005) 357--402},
  \href{https://arxiv.org/abs/nucl-ex/0505014}{{\ttfamily
  arXiv:nucl-ex/0505014}}.

\bibitem{2021ARNPS..71..377F}
L.~{Fabbietti}, V.~M. {Sarti}, and O.~V. {Doce}, ``{Study of the Strong
  Interaction Among Hadrons with Correlations at the LHC}'',
  \href{https://doi.org/10.1146/annurev-nucl-102419-034438}{{\em Annual Review
  of Nuclear and Particle Science} {\bfseries 71} (Sept., 2021) 377--402},
  \href{https://arxiv.org/abs/2012.09806}{{\ttfamily arXiv:2012.09806
  [nucl-ex]}}.

\bibitem{ALICEDetector}
{\bfseries ALICE} Collaboration, K.~Aamodt {\em et~al.}, ``{The ALICE
  experiment at the CERN LHC}'',
  \href{https://doi.org/10.1088/1748-0221/3/08/S08002}{{\em Journal of
  Instrumentation} {\bfseries 3} (2008) S08002}.
  \url{http://stacks.iop.org/1748-0221/3/i=08/a=S08002}.

\bibitem{ALICEperf}
{\bfseries ALICE} Collaboration, B.~Abelev {\em et~al.}, ``{Performance of the
  ALICE experiment at the CERN LHC}'',
  \href{https://doi.org/10.1142/S0217751X14300440}{{\em Int. J. Mod. Phys.}
  {\bfseries A29} (2014) 1430044},
  \href{https://arxiv.org/abs/1402.4476}{{\ttfamily arXiv:1402.4476
  [nucl-ex]}}.

\bibitem{ALICEVZERO}
{\bfseries ALICE} Collaboration, E.~Abbas {\em et~al.}, ``{Performance of the
  ALICE VZERO system}'',
  \href{https://doi.org/10.1088/1748-0221/8/10/P10016}{{\em JINST} {\bfseries
  8} (2013) P10016},
\href{https://arxiv.org/abs/1306.3130}{{\ttfamily arXiv:1306.3130 [nucl-ex]}}.

\bibitem{ALICEITS}
{\bfseries ALICE} Collaboration, K.~Aamodt {\em et~al.}, ``{Alignment of the
  ALICE Inner Tracking System with cosmic-ray tracks}'',
  \href{https://doi.org/10.1088/1748-0221/5/03/P03003}{{\em JINST} {\bfseries
  5} (2010) P03003},
\href{https://arxiv.org/abs/1001.0502}{{\ttfamily arXiv:1001.0502
  [physics.ins-det]}}.

\bibitem{ALICETPC}
J.~Alme, Y.~Andres, H.~Appelsh{\"a}user, S.~Bablok, N.~Bialas, {\em et~al.},
  ``{The ALICE TPC, a large 3-dimensional tracking device with fast readout for
  ultra-high multiplicity events}'',
  \href{https://doi.org/10.1016/j.nima.2010.04.042}{{\em Nucl.Instrum.Meth.}
  {\bfseries A622} (2010) 316--367},
\href{https://arxiv.org/abs/1001.1950}{{\ttfamily arXiv:1001.1950
  [physics.ins-det]}}.

\bibitem{ALICETOF}
A.~Akindinov {\em et~al.}, ``{Performance of the ALICE Time-Of-Flight detector
  at the LHC}'',
\href{https://doi.org/10.1140/epjp/i2013-13044-x}{{\em Eur. Phys. J. Plus}
  {\bfseries 128} (2013) 44}.

\bibitem{PYTHIA}
T.~Sj\"ostrand {\em et~al.}, ``{An Introduction to PYTHIA 8.2}'',
  \href{https://doi.org/10.1016/j.cpc.2015.01.024}{{\em Comput. Phys. Commun.}
  {\bfseries 191} (2015) 159--177}.

\bibitem{GEANT}
R.~Brun, F.~Bruyant, M.~Maire, A.~C. McPherson, and P.~Zanarini, {\em {GEANT 3:
  user's guide Geant 3.10, Geant 3.11; rev. version}}.
\newblock CERN, Geneva, 1987.
\newblock \url{https://cds.cern.ch/record/1119728}.

\bibitem{STAR:2003vqj}
{\bfseries STAR} Collaboration, J.~Adams {\em et~al.},
  ``{{\ensuremath{\rho^0}}} production and possible modification in {Au+Au} and
  p+p collisions at {{\ensuremath{\sqrt{S_{NN}}}} = 200 GeV}'',
  \href{https://doi.org/10.1103/PhysRevLett.92.092301}{{\em Phys. Rev. Lett.}
  {\bfseries 92} (2004) 092301},
  \href{https://arxiv.org/abs/nucl-ex/0307023}{{\ttfamily
  arXiv:nucl-ex/0307023}}.

\bibitem{ALICE:2018qdv}
{\bfseries ALICE} Collaboration, S.~Acharya {\em et~al.}, ``{Production of the
  $\rho$(770)${^{0}}$ meson in pp and Pb--Pb collisions at $\sqrt{s_{\rm NN}}$
  = 2.76 TeV}'', \href{https://doi.org/10.1103/PhysRevC.99.064901}{{\em Phys.
  Rev. C} {\bfseries 99} (2019) 064901},
  \href{https://arxiv.org/abs/1805.04365}{{\ttfamily arXiv:1805.04365
  [nucl-ex]}}.

\bibitem{OPAL:1998enc}
{\bfseries OPAL} Collaboration, K.~Ackerstaff {\em et~al.}, ``{Photon and light
  meson production in hadronic {\ensuremath{Z^{0}}} decays}'',
  \href{https://doi.org/10.1007/s100520050286}{{\em Eur. Phys. J. C} {\bfseries
  5} (1998) 411--437}, \href{https://arxiv.org/abs/hep-ex/9805011}{{\ttfamily
  arXiv:hep-ex/9805011}}.

\bibitem{DELPHI:1994qgk}
{\bfseries DELPHI} Collaboration, P.~Abreu {\em et~al.}, ``{Production
  characteristics of {\ensuremath{K^{0}}} and light meson resonances in
  hadronic decays of the {\ensuremath{Z^{0}}} }'',
  \href{https://doi.org/10.1007/BF01578668}{{\em Z. Phys. C} {\bfseries 65}
  (1995) 587--602}.

\bibitem{BEDDALL2009300}
A.~Beddall, A.~Beddall, and A.~Bingül, ``Inclusive production of the
  $\rho^{\pm}$(770) meson in hadronic decays of the z boson'',
  \href{https://doi.org/https://doi.org/10.1016/j.physletb.2008.11.018}{{\em
  Physics Letters B} {\bfseries 670} (2009) 300--306}.
  \url{https://www.sciencedirect.com/science/article/pii/S0370269308013828}.

\bibitem{PhysRev.49.519}
G.~Breit and E.~Wigner, ``Capture of slow neutrons'',
  \href{https://doi.org/10.1103/PhysRev.49.519}{{\em Phys. Rev.} {\bfseries 49}
  (Apr, 1936) 519--531}. \url{https://link.aps.org/doi/10.1103/PhysRev.49.519}.

\bibitem{PDG2024_reso}
{\bfseries Particle Data Group} Collaboration, P.~D. Group, R.~Workman, {\em
  et~al.}, ``Review of particle physics'',
  \href{https://doi.org/10.1093/ptep/ptae052}{{\em Prog. Theor. Exp. Phys.}
  {\bfseries 2024} (2024) 083C01}.
  \url{https://pdg.lbl.gov/2024/reviews/rpp2024-rev-resonances.pdf}. See the
  section on resonances.

\bibitem{PDGpionMass}
{\bfseries Particle Data Group} Collaboration, P.~D. Group, R.~Workman, {\em
  et~al.}, ``{Review of Particle Physics}'',
  \href{https://doi.org/10.1093/ptep/ptae047}{{\em Prog. Theor. Exp. Phys.}
  {\bfseries 2024} (2024) 083C01}.

\bibitem{ALICE:2018ysd}
{\bfseries ALICE} Collaboration, S.~Acharya {\em et~al.}, ``{p--p, p--$\Lambda$
  and $\Lambda$--$\Lambda$ correlations studied via femtoscopy in pp reactions
  at $\sqrt{s}$ = 7 TeV}'',
  \href{https://doi.org/10.1103/PhysRevC.99.024001}{{\em Phys. Rev. C}
  {\bfseries 99} (2019) 024001},
  \href{https://arxiv.org/abs/1805.12455}{{\ttfamily arXiv:1805.12455
  [nucl-ex]}}.

\bibitem{DelGrande:2021mju}
R.~Del~Grande, L.~\v{S}erk\v{s}nyt\.{e}, L.~Fabbietti, V.~M. Sarti, and
  D.~Mihaylov, ``{A method to remove lower order contributions in
  multi-particle femtoscopic correlation functions}'',
  \href{https://doi.org/10.1140/epjc/s10052-022-10209-z}{{\em Eur. Phys. J. C}
  {\bfseries 82} (2022) 244},
  \href{https://arxiv.org/abs/2107.10227}{{\ttfamily arXiv:2107.10227
  [nucl-th]}}.

\bibitem{HAIDENBAUER20191}
J.~Haidenbauer, ``Coupled-channel effects in hadron–hadron correlation
  functions'',
  \href{https://doi.org/https://doi.org/10.1016/j.nuclphysa.2018.10.090}{{\em
  Nuclear Physics A} {\bfseries 981} (2019) 1--16}.
  \url{https://www.sciencedirect.com/science/article/pii/S0375947418303774}.

\bibitem{PhysRev.84.1232}
E.~E. Salpeter and H.~A. Bethe, ``A relativistic equation for bound-state
  problems'', \href{https://doi.org/10.1103/PhysRev.84.1232}{{\em Phys. Rev.}
  {\bfseries 84} (Dec, 1951) 1232--1242}.
  \url{https://link.aps.org/doi/10.1103/PhysRev.84.1232}.

\bibitem{Efron:1986hys}
B.~Efron and R.~Tibshirani, ``{An introduction to the bootstrap}'', {\em
  Statist. Sci.} {\bfseries 57} (1986) 54--75.

\bibitem{Koike:1996ga}
Y.~Koike and A.~Hayashigaki, ``{QCD sum rules for {\ensuremath{\rho}},
  {\ensuremath{\omega}}, {\ensuremath{\phi correlations studied via femtoscopy
  in}} meson - nucleon scattering lengths and the mass shifts in nuclear
  medium}'', \href{https://doi.org/10.1143/PTP.98.631}{{\em Prog. Theor. Phys.}
  {\bfseries 98} (1997) 631--652},
  \href{https://arxiv.org/abs/nucl-th/9609001}{{\ttfamily
  arXiv:nucl-th/9609001}}.

\bibitem{Wang:2022zwz}
X.-Y. Wang, F.~Zeng, Q.~Wang, and L.~Zhang, ``{First extraction of the proton
  mass radius and scattering length $\vert\alpha_{\rho^{0}p}\vert$ from
  \ensuremath{\rho}$^{0}$ photoproduction}'',
  \href{https://doi.org/10.1007/s11433-022-2024-9}{{\em Sci. China Phys. Mech.
  Astron.} {\bfseries 66} (2023) 232012},
  \href{https://arxiv.org/abs/2206.09170}{{\ttfamily arXiv:2206.09170
  [nucl-th]}}.

\bibitem{Oset:2010tof}
E.~Oset and A.~Ramos, ``{Dynamically generated resonances from the vector
  octet-baryon octet interaction}'',
  \href{https://doi.org/10.1140/epja/i2010-10957-3}{{\em Eur. Phys. J. A}
  {\bfseries 44} (2010) 445--454},
  \href{https://arxiv.org/abs/0905.0973}{{\ttfamily arXiv:0905.0973 [hep-ph]}}.

\bibitem{Khemchandani:2011et}
K.~P. Khemchandani, H.~Kaneko, H.~Nagahiro, and A.~Hosaka, ``{Vector
  meson-Baryon dynamics and generation of resonances}'',
  \href{https://doi.org/10.1103/PhysRevD.83.114041}{{\em Phys. Rev. D}
  {\bfseries 83} (2011) 114041},
  \href{https://arxiv.org/abs/1104.0307}{{\ttfamily arXiv:1104.0307 [hep-ph]}}.

\bibitem{Gamermann:2011mq}
D.~Gamermann, C.~Garcia-Recio, J.~Nieves, and L.~L. Salcedo, ``{Odd Parity
  Light Baryon Resonances}'',
  \href{https://doi.org/10.1103/PhysRevD.84.056017}{{\em Phys. Rev. D}
  {\bfseries 84} (2011) 056017},
  \href{https://arxiv.org/abs/1104.2737}{{\ttfamily arXiv:1104.2737 [hep-ph]}}.

\bibitem{Garzon:2012np}
E.~J. Garzon and E.~Oset, ``{Effects of pseudoscalar-baryon channels in the
  dynamically generated vector-baryon resonances}'',
  \href{https://doi.org/10.1140/epja/i2012-12005-x}{{\em Eur. Phys. J. A}
  {\bfseries 48} (2012) 5}, \href{https://arxiv.org/abs/1201.3756}{{\ttfamily
  arXiv:1201.3756 [hep-ph]}}.

\bibitem{Vovchenko:2019pjl}
V.~Vovchenko and H.~Stoecker, ``{Thermal-FIST: A package for heavy-ion
  collisions and hadronic equation of state}'',
  \href{https://doi.org/10.1016/j.cpc.2019.06.024}{{\em Comput. Phys. Commun.}
  {\bfseries 244} (2019) 295--310},
  \href{https://arxiv.org/abs/1901.05249}{{\ttfamily arXiv:1901.05249
  [nucl-th]}}.

\bibitem{Grigoryan:2021org}
S.~Grigoryan, ``{A three component model for hadron $p_\mathrm {T}$-spectra in
  pp and Pb\textendash{}Pb collisions at the LHC}'',
  \href{https://doi.org/10.1140/epja/s10050-021-00646-x}{{\em Eur. Phys. J. A}
  {\bfseries 57} (2021) 328},
  \href{https://arxiv.org/abs/2109.07888}{{\ttfamily arXiv:2109.07888
  [hep-ph]}}.

\bibitem{ALICE:2024Benedict}
{\bfseries ALICE} Collaboration, S.~Acharya {\em et~al.}, ``{Measurement of the
  $\Sigma^+$\mbox{--}p correlation function in pp collisions at
  $\sqrt{s}=13$~TeV}'', {\em Paper} in prepration.

\end{thebibliography}\endgroup

\newpage
\appendix
\section{Calculation of production weights}
\label{appendix:ProductionWeights}
In order to calculate correlation functions in a coupled channel ansatz, the contributions from different channels have to be scaled by the production weights $\omega_j^{\rm prod.}$, describing the relative amount of pairs initially produced in channel $j$, and re-scattering to the target channel. These can be evaluated with the data-driven method introduced in Ref.~\cite{ALICE:2022yyh} and followed closely in this work. The primordial densities are estimated using the Canonical Statistical Model (CSM), with
incomplete equilibration of strangeness, as implemented in Thermal-FIST~\cite{Vovchenko:2019pjl}. The parameters used in this study are summarized in Ref.~\cite{ALICE:2022yyh}. The obtained densities are used as normalizations for the particles yields in the subsequent MC simulation of the kinematics. The latter is used to estimate the relative contribution of channel $j$ to the yield of the target channel with relative momentum $\ks<200$~\MeVc. The shape of the \pt spectra was assumed to follow a Blast-Wave distribution, where the parameters are taken from Ref.~\cite{Grigoryan:2021org} corresponding to the analyzed HM pp collisions at \thirteen data. The azimuthal angle $\varphi \in \left[0, 2\pi\right]$ and pseudorapidity $\lvert\eta\rvert<0.8$ were tuned to reproduce the geometrical acceptance of ALICE. The systematic uncertainties of the $\omega_j^{\rm prod.}$, summarized in Tab.~\ref{tab:production_weights}, is estimated by varying the upper limit of the \ks threshold of the produced pairs by $\pm 100$~\MeVc and the parameters of the CSM within $1\sigma$.

\begin{table}[ht]
\begin{center}
\caption{Values of the production weights $\omega_j^{\rm prod.}$ for the genuine \phiP and \rhop correlations. All values have an uncertainty of $\pm 20\%$.}

\begin{tabular}{c|cc}
\hline\hline
channel$-j$ & $\omega_j^{prod}$ (\rhop) & $\omega_j^{prod}$  (\phiP)   \\ [2mm] \hline
$\uprho^0 p$ & $1$ & $6.24$  \\ 
 $\uprho^+ n$  & $0.95$ & $5.94$  \\ 
$\upomega p$ & $0.92$ & $5.77$ \\ 
 $\upphi p$ & $0.16$ &  $1$  \\ 
 $K^{*+}\Lambda$ & $0.10$  & $0.65$  \\ 
$K^{*o}\Sigma^+$ & $0.067$  &  $0.41$  \\ 
 $K^{*+}\Sigma^0$  & $0.069$  &  $0.43$  \\ [0mm] 
\hline\hline
\end{tabular}
\label{tab:production_weights}
\end{center}
\end{table}

\section{Constraining the source}
\label{appendix:Source}
The Resonance Source Model (RSM), introduced in Ref.~\cite{ALICE:2020ibs} for baryon--baryon and extended to meson--baryon and meson--meson systems in Refs.~\cite{ALICE:2023sjd, ALICE:2024Marcel}, demonstrates a universal common hadron source for small collision systems, provided that the particle-specific feed-down from strongly decaying resonances is properly accounted for. The primordial and strong feed down composition for each hadron are estimated using the same CSM setup used for the determination of the $\omega_j^{\rm prod.}$. For each particle the mass and lifetime of each strongly decaying resonance are embedded in two effective parameters \masseff and \ctaueff, which are the average mass and decay length computed using the abundances of the considered resonances as weights. The strongly decaying resonances are explicitly considered for p (and n)~\cite{ALICE:2020ibs}, $\Lambda$~\cite{ALICE:2020ibs}, and  $\Sigma^{+/-}$~\cite{ALICE:2022yyh, ALICE:2024Benedict} using previous works. For $\phi$ a 100\% primordial fraction is assumed~\cite{ALICE:2019hdt}. Finally for $\uprho^{0/+}$, $\upomega(782)$, and $\rm{K}^{*+/0}$ new summary parameters according to Ref.~\cite{ALICE:2020ibs} are estimated and presented in Tab.~\ref{tab:RSMparameters}. A systematic uncertainty of $\pm 10\%$ is considered for the RSM calculation, accounting for any systematic effects related to the hadronic cocktail obtained from the CSM calculation as well as the averaging procedure.

\begin{table}[ht]
\begin{center}
\caption{Weight parameters used for the decomposition of the \rhop correlation function. All values have an uncertainty of $\pm 10\%$.}

\begin{tabular}{l | c | c | c}
\hline\hline
hadron & \masseff (\GeVmass) & \ctaueff (fm) & primordial [\%]\\

\hline 
$\uprho^{0/+}$ & 1.349 & 1.3 & 72.2\\
$\upomega(782)$  & 1.331 & 1.3 & 67.7\\
$\rm{K}^{*+/0}$ & 1.397 & 1.6 & 68.9\\
\hline \hline
\end{tabular}
\label{tab:RSMparameters}
\end{center}
\end{table}

\section{Details of model parameters}
\label{appendix:SubtractionConstants}
The subtraction constants (SCs) $a_l$ are part of the loop function replacing the divergence for a given dimensional regularization scale $\mu$, which is taken to be $630$~MeV in this work, see Ref.~\cite{Feijoo:2024bvn}. Their values are sensitive to the pole position and pole composition of the dynamically generated states in the employed unitarised \(\chi\)EFT. The obtained values for the $5$ SCs and of the normalization for the \rhop correlation function are shown in Tab.~\ref{tab:outputs_fits}.

\begin{table}[ht]
\centering
\caption{Values of the model parameters obtained from the bootstrap, together with the natural values.} 
\begin{tabular}{lcc}
\hline\hline \\[-2.5mm]
     & \textbf{Pure theoretical}  &  \textbf{Bootstrap}      \\                             
\hline \\[-2.5mm]
$a_{\uprho N}$    & $-2$  \, (fixed)   &  $-2$  (fixed)  \\
$a_{\upomega N}$   &  $-2$ \, (fixed)   & $ -3.04 \pm 0.73$       \\
$a_{\upphi N}$     &  $-2$ \, (fixed)   &  $ -3.15 \pm 0.37$       \\
$a_{K^{*}\Lambda}$ &  $-2$ \, (fixed) &  $ -1.98 \pm 0.08$      \\
$a_{K^{*}\Sigma}$   &  $-2$ \, (fixed) &  $ -1.95 \pm 0.08$      \\
$N_{D}$           & $1$  \, (fixed)   &  $ 0.988\pm 0.004 $  \\
\hline\hline
\end{tabular}
\label{tab:outputs_fits}  
\end{table}

\section{The ALICE Collaboration}
\label{app:collab}
\begin{flushleft} 
\small

I.J.~Abualrob\,\orcidlink{0009-0005-3519-5631}\,$^{\rm 114}$, 
S.~Acharya\,\orcidlink{0000-0002-9213-5329}\,$^{\rm 50}$, 
G.~Aglieri Rinella\,\orcidlink{0000-0002-9611-3696}\,$^{\rm 32}$, 
L.~Aglietta\,\orcidlink{0009-0003-0763-6802}\,$^{\rm 24}$, 
M.~Agnello\,\orcidlink{0000-0002-0760-5075}\,$^{\rm 29}$, 
N.~Agrawal\,\orcidlink{0000-0003-0348-9836}\,$^{\rm 25}$, 
Z.~Ahammed\,\orcidlink{0000-0001-5241-7412}\,$^{\rm 133}$, 
S.~Ahmad\,\orcidlink{0000-0003-0497-5705}\,$^{\rm 15}$, 
I.~Ahuja\,\orcidlink{0000-0002-4417-1392}\,$^{\rm 36}$, 
ZUL.~Akbar$^{\rm 81}$, 
A.~Akindinov\,\orcidlink{0000-0002-7388-3022}\,$^{\rm 139}$
% ...

\section*{Affiliation Notes}

$^{\rm I}$ Deceased\\
$^{\rm II}$ Also at: Max-Planck-Institut fur Physik, Munich, Germany\\
$^{\rm III}$ Also at: Czech Technical University in Prague (CZ)\\
$^{\rm IV}$ Also at: Italian National Agency for New Technologies, Energy and Sustainable Economic Development (ENEA), Bologna, Italy\\
$^{\rm V}$ Also at: Instituto de Fisica da Universidade de Sao Paulo\\
$^{\rm VI}$ Also at: Dipartimento DET del Politecnico di Torino, Turin, Italy\\
$^{\rm VII}$ Also at: Department of Applied Physics, Aligarh Muslim University, Aligarh, India\\
$^{\rm VIII}$ Also at: Institute of Theoretical Physics, University of Wroclaw, Poland\\
$^{\rm IX}$ Also at: Facultad de Ciencias, Universidad Nacional Aut\'{o}noma de M\'{e}xico, Mexico City, Mexico\\

\section*{Collaboration Institutes}

$^{1}$ A.I. Alikhanyan National Science Laboratory (Yerevan Physics Institute) Foundation, Yerevan, Armenia\\
$^{2}$ AGH University of Krakow, Cracow, Poland\\
$^{3}$ Bogolyubov Institute for Theoretical Physics, National Academy of Sciences of Ukraine, Kyiv, Ukraine\\
$^{4}$ Bose Institute, Department of Physics  and Centre for Astroparticle Physics and Space Science (CAPSS), Kolkata, India\\
$^{5}$ California Polytechnic State University, San Luis Obispo, California, United States\\
$^{6}$ Central China Normal University, Wuhan, China\\
$^{7}$ Centro de Aplicaciones Tecnol\'{o}gicas y Desarrollo Nuclear (CEADEN), Havana, Cuba\\
$^{8}$ Centro de Investigaci\'{o}n y de Estudios Avanzados (CINVESTAV), Mexico City and M\'{e}rida, Mexico\\
$^{9}$ Chicago State University, Chicago, Illinois, United States\\
$^{10}$ China Nuclear Data Center, China Institute of Atomic Energy, Beijing, China\\
$^{11}$ China University of Geosciences, Wuhan, China\\
$^{12}$ Chungbuk National University, Cheongju, Republic of Korea\\
$^{13}$ Comenius University Bratislava, Faculty of Mathematics, Physics and Informatics, Bratislava, Slovak Republic\\
$^{14}$ Creighton University, Omaha, Nebraska, United States\\
$^{15}$ Department of Physics, Aligarh Muslim University, Aligarh, India\\
$^{16}$ Department of Physics, Pusan National University, Pusan, Republic of Korea\\
$^{17}$ Department of Physics, Sejong University, Seoul, Republic of Korea\\
$^{18}$ Department of Physics, University of California, Berkeley, California, United States\\
$^{19}$ Department of Physics, University of Oslo, Oslo, Norway\\
$^{20}$ Department of Physics and Technology, University of Bergen, Bergen, Norway\\
$^{21}$ Dipartimento di Fisica, Universit\`{a} di Pavia, Pavia, Italy\\
$^{22}$ Dipartimento di Fisica dell'Universit\`{a} and Sezione INFN, Cagliari, Italy\\
$^{23}$ Dipartimento di Fisica dell'Universit\`{a} and Sezione INFN, Trieste, Italy\\
$^{24}$ Dipartimento di Fisica dell'Universit\`{a} and Sezione INFN, Turin, Italy\\
$^{25}$ Dipartimento di Fisica e Astronomia dell'Universit\`{a} and Sezione INFN, Bologna, Italy\\
$^{26}$ Dipartimento di Fisica e Astronomia dell'Universit\`{a} and Sezione INFN, Catania, Italy\\
$^{27}$ Dipartimento di Fisica e Astronomia dell'Universit\`{a} and Sezione INFN, Padova, Italy\\
$^{28}$ Dipartimento di Fisica `E.R.~Caianiello' dell'Universit\`{a} and Gruppo Collegato INFN, Salerno, Italy\\
$^{29}$ Dipartimento DISAT del Politecnico and Sezione INFN, Turin, Italy\\
$^{30}$ Dipartimento di Scienze MIFT, Universit\`{a} di Messina, Messina, Italy\\
$^{31}$ Dipartimento Interateneo di Fisica `M.~Merlin' and Sezione INFN, Bari, Italy\\
$^{32}$ European Organization for Nuclear Research (CERN), Geneva, Switzerland\\
$^{33}$ Faculty of Electrical Engineering, Mechanical Engineering and Naval Architecture, University of Split, Split, Croatia\\
$^{34}$ Faculty of Nuclear Sciences and Physical Engineering, Czech Technical University in Prague, Prague, Czech Republic\\
$^{35}$ Faculty of Physics, Sofia University, Sofia, Bulgaria\\
$^{36}$ Faculty of Science, P.J.~\v{S}af\'{a}rik University, Ko\v{s}ice, Slovak Republic\\
$^{37}$ Faculty of Technology, Environmental and Social Sciences, Bergen, Norway\\
$^{38}$ Frankfurt Institute for Advanced Studies, Johann Wolfgang Goethe-Universit\"{a}t Frankfurt, Frankfurt, Germany\\
$^{39}$ Fudan University, Shanghai, China\\
$^{40}$ Gangneung-Wonju National University, Gangneung, Republic of Korea\\
$^{41}$ Gauhati University, Department of Physics, Guwahati, India\\
$^{42}$ Helmholtz-Institut f\"{u}r Strahlen- und Kernphysik, Rheinische Friedrich-Wilhelms-Universit\"{a}t Bonn, Bonn, Germany\\
$^{43}$ Helsinki Institute of Physics (HIP), Helsinki, Finland\\
$^{44}$ High Energy Physics Group,  Universidad Aut\'{o}noma de Puebla, Puebla, Mexico\\
$^{45}$ Horia Hulubei National Institute of Physics and Nuclear Engineering, Bucharest, Romania\\
$^{46}$ HUN-REN Wigner Research Centre for Physics, Budapest, Hungary\\
$^{47}$ Indian Institute of Technology Bombay (IIT), Mumbai, India\\
$^{48}$ Indian Institute of Technology Indore, Indore, India\\
$^{49}$ INFN, Laboratori Nazionali di Frascati, Frascati, Italy\\
$^{50}$ INFN, Sezione di Bari, Bari, Italy\\
$^{51}$ INFN, Sezione di Bologna, Bologna, Italy\\
$^{52}$ INFN, Sezione di Cagliari, Cagliari, Italy\\
$^{53}$ INFN, Sezione di Catania, Catania, Italy\\
$^{54}$ INFN, Sezione di Padova, Padova, Italy\\
$^{55}$ INFN, Sezione di Pavia, Pavia, Italy\\
$^{56}$ INFN, Sezione di Torino, Turin, Italy\\
$^{57}$ INFN, Sezione di Trieste, Trieste, Italy\\
$^{58}$ Inha University, Incheon, Republic of Korea\\
$^{59}$ Institute for Gravitational and Subatomic Physics (GRASP), Utrecht University/Nikhef, Utrecht, Netherlands\\
$^{60}$ Institute of Experimental Physics, Slovak Academy of Sciences, Ko\v{s}ice, Slovak Republic\\
$^{61}$ Institute of Physics, Homi Bhabha National Institute, Bhubaneswar, India\\
$^{62}$ Institute of Physics of the Czech Academy of Sciences, Prague, Czech Republic\\
$^{63}$ Institute of Space Science (ISS), Bucharest, Romania\\
$^{64}$ Institut f\"{u}r Kernphysik, Johann Wolfgang Goethe-Universit\"{a}t Frankfurt, Frankfurt, Germany\\
$^{65}$ Instituto de Ciencias Nucleares, Universidad Nacional Aut\'{o}noma de M\'{e}xico, Mexico City, Mexico\\
$^{66}$ Instituto de F\'{i}sica, Universidade Federal do Rio Grande do Sul (UFRGS), Porto Alegre, Brazil\\
$^{67}$ Instituto de F\'{\i}sica, Universidad Nacional Aut\'{o}noma de M\'{e}xico, Mexico City, Mexico\\
$^{68}$ iThemba LABS, National Research Foundation, Somerset West, South Africa\\
$^{69}$ Jeonbuk National University, Jeonju, Republic of Korea\\
$^{70}$ Johann-Wolfgang-Goethe Universit\"{a}t Frankfurt Institut f\"{u}r Informatik, Fachbereich Informatik und Mathematik, Frankfurt, Germany\\
$^{71}$ Korea Institute of Science and Technology Information, Daejeon, Republic of Korea\\
$^{72}$ Laboratoire de Physique Subatomique et de Cosmologie, Universit\'{e} Grenoble-Alpes, CNRS-IN2P3, Grenoble, France\\
$^{73}$ Lawrence Berkeley National Laboratory, Berkeley, California, United States\\
$^{74}$ Lund University Department of Physics, Division of Particle Physics, Lund, Sweden\\
$^{75}$ Nagasaki Institute of Applied Science, Nagasaki, Japan\\
$^{76}$ Nara Women{'}s University (NWU), Nara, Japan\\
$^{77}$ National and Kapodistrian University of Athens, School of Science, Department of Physics , Athens, Greece\\
$^{78}$ National Centre for Nuclear Research, Warsaw, Poland\\
$^{79}$ National Institute of Science Education and Research, Homi Bhabha National Institute, Jatni, India\\
$^{80}$ National Nuclear Research Center, Baku, Azerbaijan\\
$^{81}$ National Research and Innovation Agency - BRIN, Jakarta, Indonesia\\
$^{82}$ Niels Bohr Institute, University of Copenhagen, Copenhagen, Denmark\\
$^{83}$ Nikhef, National institute for subatomic physics, Amsterdam, Netherlands\\
$^{84}$ Nuclear Physics Group, STFC Daresbury Laboratory, Daresbury, United Kingdom\\
$^{85}$ Nuclear Physics Institute of the Czech Academy of Sciences, Husinec-\v{R}e\v{z}, Czech Republic\\
$^{86}$ Oak Ridge National Laboratory, Oak Ridge, Tennessee, United States\\
$^{87}$ Ohio State University, Columbus, Ohio, United States\\
$^{88}$ Physics department, Faculty of science, University of Zagreb, Zagreb, Croatia\\
$^{89}$ Physics Department, Panjab University, Chandigarh, India\\
$^{90}$ Physics Department, University of Jammu, Jammu, India\\
$^{91}$ Physics Program and International Institute for Sustainability with Knotted Chiral Meta Matter (WPI-SKCM$^{2}$), Hiroshima University, Hiroshima, Japan\\
$^{92}$ Physikalisches Institut, Eberhard-Karls-Universit\"{a}t T\"{u}bingen, T\"{u}bingen, Germany\\
$^{93}$ Physikalisches Institut, Ruprecht-Karls-Universit\"{a}t Heidelberg, Heidelberg, Germany\\
$^{94}$ Physik Department, Technische Universit\"{a}t M\"{u}nchen, Munich, Germany\\
$^{95}$ Politecnico di Bari and Sezione INFN, Bari, Italy\\
$^{96}$ Research Division and ExtreMe Matter Institute EMMI, GSI Helmholtzzentrum f\"ur Schwerionenforschung GmbH, Darmstadt, Germany\\
$^{97}$ Saga University, Saga, Japan\\
$^{98}$ Saha Institute of Nuclear Physics, Homi Bhabha National Institute, Kolkata, India\\
$^{99}$ School of Physics and Astronomy, University of Birmingham, Birmingham, United Kingdom\\
$^{100}$ Secci\'{o}n F\'{\i}sica, Departamento de Ciencias, Pontificia Universidad Cat\'{o}lica del Per\'{u}, Lima, Peru\\
$^{101}$ Stefan Meyer Institut f\"{u}r Subatomare Physik (SMI), Vienna, Austria\\
$^{102}$ SUBATECH, IMT Atlantique, Nantes Universit\'{e}, CNRS-IN2P3, Nantes, France\\
$^{103}$ Sungkyunkwan University, Suwon City, Republic of Korea\\
$^{104}$ Suranaree University of Technology, Nakhon Ratchasima, Thailand\\
$^{105}$ Technical University of Ko\v{s}ice, Ko\v{s}ice, Slovak Republic\\
$^{106}$ The Henryk Niewodniczanski Institute of Nuclear Physics, Polish Academy of Sciences, Cracow, Poland\\
$^{107}$ The University of Texas at Austin, Austin, Texas, United States\\
$^{108}$ Universidad Aut\'{o}noma de Sinaloa, Culiac\'{a}n, Mexico\\
$^{109}$ Universidade de S\~{a}o Paulo (USP), S\~{a}o Paulo, Brazil\\
$^{110}$ Universidade Estadual de Campinas (UNICAMP), Campinas, Brazil\\
$^{111}$ Universidade Federal do ABC, Santo Andre, Brazil\\
$^{112}$ Universitatea Nationala de Stiinta si Tehnologie Politehnica Bucuresti, Bucharest, Romania\\
$^{113}$ University of Derby, Derby, United Kingdom\\
$^{114}$ University of Houston, Houston, Texas, United States\\
$^{115}$ University of Jyv\"{a}skyl\"{a}, Jyv\"{a}skyl\"{a}, Finland\\
$^{116}$ University of Kansas, Lawrence, Kansas, United States\\
$^{117}$ University of Liverpool, Liverpool, United Kingdom\\
$^{118}$ University of Science and Technology of China, Hefei, China\\
$^{119}$ University of South-Eastern Norway, Kongsberg, Norway\\
$^{120}$ University of Tennessee, Knoxville, Tennessee, United States\\
$^{121}$ University of the Witwatersrand, Johannesburg, South Africa\\
$^{122}$ University of Tokyo, Tokyo, Japan\\
$^{123}$ University of Tsukuba, Tsukuba, Japan\\
$^{124}$ Universit\"{a}t M\"{u}nster, Institut f\"{u}r Kernphysik, M\"{u}nster, Germany\\
$^{125}$ Universit\'{e} Clermont Auvergne, CNRS/IN2P3, LPC, Clermont-Ferrand, France\\
$^{126}$ Universit\'{e} de Lyon, CNRS/IN2P3, Institut de Physique des 2 Infinis de Lyon, Lyon, France\\
$^{127}$ Universit\'{e} de Strasbourg, CNRS, IPHC UMR 7178, F-67000 Strasbourg, France, Strasbourg, France\\
$^{128}$ Universit\'{e} Paris-Saclay, Centre d'Etudes de Saclay (CEA), IRFU, D\'{e}partment de Physique Nucl\'{e}aire (DPhN), Saclay, France\\
$^{129}$ Universit\'{e}  Paris-Saclay, CNRS/IN2P3, IJCLab, Orsay, France\\
$^{130}$ Universit\`{a} degli Studi di Foggia, Foggia, Italy\\
$^{131}$ Universit\`{a} del Piemonte Orientale, Vercelli, Italy\\
$^{132}$ Universit\`{a} di Brescia, Brescia, Italy\\
$^{133}$ Variable Energy Cyclotron Centre, Homi Bhabha National Institute, Kolkata, India\\
$^{134}$ Warsaw University of Technology, Warsaw, Poland\\
$^{135}$ Wayne State University, Detroit, Michigan, United States\\
$^{136}$ Yale University, New Haven, Connecticut, United States\\
$^{137}$ Yildiz Technical University, Istanbul, Turkey\\
$^{138}$ Yonsei University, Seoul, Republic of Korea\\
$^{139}$ Affiliated with an institute formerly covered by a cooperation agreement with CERN\\
$^{140}$ Affiliated with an international laboratory covered by a cooperation agreement with CERN.\\

\end{flushleft} 
  
\end{document}